\documentclass[%
 aip,
 jmp,%
 amsmath,amssymb,
preprint,%
]{revtex4-1}

\usepackage{graphicx}
\usepackage{dcolumn}
\usepackage{bm}
\usepackage{xcolor}
 \usepackage[version=3]{mhchem}
\usepackage{setspace}
\usepackage{threeparttable}
\usepackage{hyperref}

\begin{document}

\preprint{Version \today}

\title[]{Multiconfigurational Short-Range Density-Functional Theory for Open-Shell Systems}

\author{Erik Donovan Hedeg{\aa}rd}
\email{erik.hedegard@teokem.lu.se}
\affiliation{Department of Theoretical Chemistry, Lund University, Kemicentrum P.O.~Box 124, SE-221 00 Lund, Sweden}
\author{Julien Toulouse}
\affiliation{Laboratoire de Chimie Th{\'e}orique, Sorbonne Universit\'e and CNRS, Paris, France}
\author{Hans J{\o}rgen Aagaard Jensen}
\email{hjj@sdu.dk}
\affiliation{Department of Physics, Chemistry and Pharmacy, University of Southern Denmark, Odense M, Denmark}


\begin{abstract}
Many chemical systems cannot be described by quantum chemistry methods based on a single-reference wave function. Accurate predictions of energetic and spectroscopic properties require a delicate balance between describing the most important configurations (static correlation) and obtaining dynamical correlation efficiently. The former is most naturally done through a multiconfigurational wave function, whereas the latter can be done by e.g.~perturbation theory. We have employed a different strategy, namely a hybrid between multiconfigurational wave functions and density-functional theory (DFT) based on range separation. The method is denoted multiconfigurational short-range DFT (MC--srDFT) and is more efficient than perturbative approaches as it capitalizes on the efficient treatment of the (short-range) dynamical correlation by DFT approximations. In turn, the method also improves DFT with standard approximations through the ability of multiconfigurational wave functions to recover large parts of the static correlation. Until now, our implementation was restricted to closed-shell systems and to lift this restriction, we present here the generalization of MC--srDFT to open-shell cases. The additional terms required to treat open-shell systems are derived and implemented in the {\sc DALTON} program. This new method for open-shell systems is illustrated on dioxygen and \ce{[Fe(H2O)6]^{3+}}.  
%
\end{abstract}

\keywords{range-separated DFT, MCSCF, open shells}
\maketitle

\section{\label{intro} Introduction}

Quantum chemistry (QC) methods have become paramount to gain insight in experimental investigations 
and to support findings from experimental studies.
Moreover, QC methods can be employed to predict novel molecular systems with a given desired property. 
Naturally, the above uses of theoretical methods require that these methods are both efficient and accurate, also in cases where reference data 
might not exist or be inconclusive. Density-functional theory (DFT) has to a large extent fulfilled these requirements\cite{burke2012}, 
and can for many systems be employed routinely. Still, DFT with standard approximations is known to fail for systems that cannot be described by a single-reference wave function.\cite{cohen2008,cohen2012} Such systems are often characterized by dense orbital manifolds and 
several (near)-degenerate states. In wave-function theory (WFT), such systems are best described with a 
multiconfigurational wave function.\cite{szalay2012} 
These exist in a number of different forms of which many have in common the definition of a \textit{complete active space} CAS($m,n$) 
of $m$ electrons in $n$ orbitals. In this active space, all configurations (fulfilling additional spin and symmetry constraints) are included.
This is usually combined with optimization of orbital parameters in what is denoted a complete-active-space self-consistent field (CASSCF) procedure. Unfortunately, 
the computational demands increase dramatically with the size of the active space, 
and many systems require
active spaces beyond the current limitations to give physically meaningful results.
In recent years, several groups have focused on lifting the limitations 
for the size of active space with methods such as the 
density-matrix renormalization group (DMRG)\cite{marti2010b,chan2011,kurashige2014,knecht2016}, 
quantum Monte Carlo (QMC)\cite{filippi2016,manni2016,thomas2015}, restricted-active-space (RAS)\cite{olsen1988} or  
generalized-active-space (GAS) methods.\cite{FleOlsMar2001,ThyFleJen2008,ma2011}
Yet, even with extended active orbital spaces, essential parts of the
remaining dynamical electron correlation cannot be obtained, 
except for the smallest systems where the number of occupied valence orbitals is small. 
Typically, dynamical correlation for such multireference systems is obtained \textit{after} initially  
obtaining a correct representation of a zeroth-order Hamiltonian (including static correlation). 
The exact nature of the subsequent steps responsible for recovering 
dynamical correlation depends on the chosen method, but well-known examples are multireference perturbation theories\cite{andersson1990,andersson1992,RASPT2,GASPT2,angeli2001} such as
complete-active-space second-order perturbation theory (CASPT2)\cite{andersson1990,andersson1992} and  
\textit{n}-electron valence state perturbation theory (NEVPT2)\cite{angeli2001}.

We have in a number of recent papers employed a different strategy, namely a hybrid between 
MCSCF and DFT that capitalizes on the fairly efficient treatment 
of the (short-range) dynamical correlation by semi-local DFT approximations and 
the ability of CASSCF or more general MCSCF models to recover large parts of the static correlation. 
 A number of different CAS-DFT hybrid methods have been suggested\cite{grimme1999,marian2008,manni2014,gagliardi2017,savin1995,leininger1997,fromager2007,ShaSavJenTou-JCP-12,fromager2013,hedegaard2013b},
many of which are under active development. There is one important difference between these methods: some add DFT after initial optimization of the multiconfigurational wave function\cite{manni2014,gagliardi2017}, whereas others optimize the DFT and wave function parts simultaneously\cite{grimme1999,marian2008,savin1995,leininger1997,fromager2007,ShaSavJenTou-JCP-12,fromager2013,hedegaard2013b}. The method reported here belongs to the latter kind.  A challenge for all CAS-DFT hybrids is to avoid double counting of electron correlation, as the CAS wave functions will invariably include some of the dynamical correlation.  
The method that we discuss in this paper, and that we have been developing since several years \cite{fromager2007,fromager2013,hedegaard2013b,hedegaard2015a,hedegaard2015b,hedegaard2016a,hedegaard2016b,hubert2016a,hubert2016b}, stringently avoids this double counting by means of a range separation of the two-electron repulsion operator\cite{savinbook,savin1995}. 
The \textit{long-range} part is then described by WFT,
whereas the \textit{short-range} part is described by a tailored short-range DFT functional.\cite{toulouse2004a}
When employing an MC(SCF) wave function as the 
long-range component, we have dubbed this method MC--srDFT.
For the common choice of CAS for the MCSCF part, this becomes CAS--srDFT.
However, it should be emphasized, that the method of range separation also allows for other wave-function 
\textit{ans{\"a}tze}: CI--srDFT\cite{leininger1997,pedersenphd2004}, MP2--srDFT\cite{angyan2005,fromager2008}, RPA-srDFT\cite{TouGerJanSavAng-PRL-09,JanHenScu-JCP-09}, CC--srDFT\cite{goll2005}, RAS--srDFT, NEVPT2--srDFT\cite{fromager2010}, and DMRG--srDFT\cite{hedegaard2015b} methods have all been implemented.    

Our MC--srDFT implementation in a development version of {\sc DALTON}\cite{DALTON2016,WCMS:WCMS1172}
employs a general implementation for the long-range MCSCF wave function that allows any spin multiplicity.
However, the implementation has until now been restricted to short-range functionals depending only on the total density and not on the spin density.
Calculations have accordingly been done on closed-shell systems only.
It should in this regard be noted that introducing a dependence on the spin density in the DFT part is not necessary from a theoretical point of view where the energy is solely defined from the total density.
Yet, approximate semi-local functionals must depend on the spin density to be accurate for open-shell systems,
and a semi-local srDFT implementation based solely on the total density in practice excludes useful calculations on many open-shell target systems where we expect MC--srDFT to be able to perform well.
For example, many transition metals form open-shell complexes, and these are notorious for displaying multireference character.

Since the original formulation of MC--srDFT\cite{pedersenphd2004},
several srDFT functionals have been generalized to include the spin density.\cite{paziani2006,goll2006}
In this paper, we extend the MC--srDFT method with these spin-dependent functionals, both for a local-density approximation (LDA) and a generalized-gradient approximation (GGA). 
This enables meaningful calculations for open-shell systems, which in particular will be beneficial for our endeavors
to describe transition-metal complexes with CAS--srDFT or GAS--srDFT as a computationally cheaper alternative to CASPT2, RASPT2, and NEVPT2.

As a first test of our novel implementation we investigate two prototypical open-shell molecules:
the \ce{O2} molecule and the transition metal complex \ce{[Fe(H2O)6]^{3+}}.
We focus here on rather simple systems which allow for a thorough investigation, and where literature data are available.
For both the $^{3}\Sigma^{-}_{g}$ ground-state of \ce{O2} and the spin-state splitting in \ce{[Fe(H2O)6]^{3+}} we also carry out a 
small study on the dependence of CAS--srDFT on the range-separation parameter $\mu$, similarly to what has previously been done for a number 
of small, closed-shell p-block molecules and atoms.\cite{fromager2007} 
It should be emphasized that although \ce{O2} is a prototypical open-shell system for which a wealth of literature data 
is available\cite{schweitzer2003}, the spin-state splitting between the  $^{3}\Sigma^{-}_{g}$ ground-state and the first singlet state, 
$^{1}\Delta_{g}$, is surprisingly demanding in terms of the theoretical treatment.\cite{klotzchristel1984,gadzhiev2013,zen2014} Likewise, 
the spin-state splitting between the lowest electronic states, $^{6}\text{A}_{g}$ and $^{4}\text{T}_{1g}$ (assuming idealized, octahedral symmetry) 
in \ce{[Fe(H2O)6]^{3+}} has turned out to be challenging; 
there is currently no consensus on which method is the most accurate, as both CASPT2 and CCSD(T) have been shown to overestimate the spin-state splitting somewhat.\cite{ghosh2003}
Accordingly, both \ce{O2} and \ce{[Fe(H2O)6]^{3+}} have low-lying excited states with different spin multiplicity than the ground-state,
and for both a high-level method is evidently needed to obtain correct results.

The paper is organized as follows. In Section \ref{SEC_LR_MC_SRDFT} we describe the necessary theory for the extension of 
MC--srDFT to srDFT functionals that include the spin density as well as the gradient of the spin density,
and in Section \ref{implement} we describe implementation details.
The theory and implementation details for the closed-shell singlet case is obviously a special case, which previously only has been published in the PhD dissertation by J. K. Pedersen.\cite{pedersenphd2004}
Next, we provide the computational details that were employed for the test
calculations (Section \ref{comp}), and in Section \ref{results} we discuss the results. Finally, we give a brief conclusion and outlook in Section \ref{conclusion}. 

\section{Range-Separated Multiconfigurational Density Functional Theory}\label{SEC_LR_MC_SRDFT}
 The MC--srDFT method employs a range separation of the two-electron operator\cite{savin1995,savinbook}
\begin{align}
 \hat{g}(1,2) =  \hat{g}^{\text{lr}}(1,2) +  \hat{g}^{\text{sr}}(1,2),  \label{two-elec-operator}
\end{align}
in which the exact definition of $\hat{g}^{\text{lr}}(1,2)$ and $\hat{g}^{\text{sr}}(1,2)$ can differ\cite{toulouse2004b}.
In this work we exclusively use the error function for the range separation:
\begin{align}
 \hat{g}^{\text{lr}}(1,2) = \frac{\text{erf} (\mu r_{12})}{r_{12}}   \quad\text{and}\quad
 \hat{g}^{\text{sr}}(1,2) = \frac{1- \text{erf} (\mu r_{12})}{r_{12}}  , \label{lrsrpart}
\end{align}
where $r_{12} = | \mathbf{r}_{1} - \mathbf{r}_{2} |$ and $\mu$ is the range-separation parameter, given in reciprocal bohr.  
Eq.~\eqref{two-elec-operator} is given in atomic units which we will employ throughout this paper.
The energy of a range-separated WFT-DFT hybrid is then given by
\begin{align}
 E(\bm{\lambda}) & = E_{\text{lr}}(\bm{\lambda}) + E_{\text{sr-H}}[\rho_{C} ] + E_{\text{sr-xc}}[\rho_{C},\rho_S] \notag \\
& = \langle\Psi^{\text{lr}}(\bm{\lambda})\vert \hat{H}^{\text{lr}}\vert\Psi^{\text{lr}}(\bm{\lambda})\rangle 
+ E_{\text{sr-H}}[\rho_{C} ] + E_{\text{sr-xc}}[\rho_{C} ,\rho_{S}] , 
\label{E_MCsrDFT}
\end{align}
where $\bm{\lambda}$ are the wave-function parameters, to be defined below. 
We denote the electron charge density as $\rho_{C}$ and the spin density as $\rho_{S}$,
and in order not to overload the notation we let it be implicit that they depend on the wave-function parameters $\bm{\lambda}$:
$\rho_{C}\equiv\rho_{C}(\mathbf{r},\bm{\lambda})$ and $\rho_{S}\equiv\rho_{S}(\mathbf{r},\bm{\lambda})$.  We define the short-range Hartree and exchange-correlation functionals, $E_{\text{sr-H}}[\rho_{C} ]$ and $E_{\text{sr-xc}}[\rho_{C} ,\rho_{S}]$, below after first defining the quantities involved in the long-range part of the energy in Eq.~\eqref{E_MCsrDFT}. The long-range Hamiltonian $\hat{H}^{\text{lr}}$  takes a form similar to the regular electronic (non- or scalar-relativistic) Hamiltonian in second quantization
\begin{equation}
 \hat{H}^{\text{lr}} = \sum_{pq}h_{pq}\hat{E}_{pq} + \frac{1}{2}\sum_{pqrs} g^{\rm lr}_{pq,rs}\hat{e}_{pq,rs} + V_{\text{nn}} \label{hamilton} , 
\end{equation}
with the standard one-electron integrals $h_{pq}$ and nuclei-nuclei interaction energy $V_{\text{nn}}$, but with the two-electron Coulomb integrals replaced by integrals with the modified long-range interaction in Eq.~\eqref{lrsrpart}, 
\begin{align}
g^{\text{lr}}_{pq,rs} & = \langle \phi_{p} \phi_{r} \vert \hat{g}^{\text{lr}}(1,2) \vert \phi_{q} \phi_{s}\rangle, \label{g_pqrs}
\end{align} 
in an orthonormal basis of spatial orbitals $\{\phi_p\}$.
The second-quantized operators in Eq.~\eqref{hamilton} have their usual meaning: $\hat{E}_{pq}=\hat{a}_{p\alpha}^\dagger\hat{a}_{q\alpha} + \hat{a}_{p\beta}^\dagger\hat{a}_{q\beta}$ is the singlet excitation operator and $\hat{e}_{pq,rs} = \hat{E}_{pq} \hat{E}_{rs} - \delta_{qr} \hat{E}_{ps}$. The indices $p,q,r,s$ denote general orbitals.  
For the MC(SCF) wave function parameters we use the notation $\bm{\lambda}^T = (\bm{d}^T,\bm{\kappa}^T)$ where the row vectors
$\bm{d}^T=\{ d_j \}$ and $\bm{\kappa}^T = \{\kappa_{pq} \}$ designate
the configuration coefficients and the orbital rotation parameters, respectively.
The real-valued MC wave function is parameterized as
\begin{equation}
 \vert \Psi^{\text{lr}}(\bm{\lambda})\rangle =  e^{-\hat{\kappa}} \left(
\frac{\vert 0 \rangle + \hat{\mathcal{P}}\vert\, \bm{d}\, \rangle}{\sqrt{1 + \langle\, \bm{d}\, \vert\mathcal{\hat{P}}\vert\, \bm{d}\, \rangle}} 
\right) , 
\label{srdftwavefunc}
\end{equation}
where $\vert 0\rangle$ denotes a normalized reference state 
\begin{align}
 \vert 0\rangle = \sum_{j}c_{j}\vert j\rangle, 
\end{align}
while  
\begin{equation}
 \vert\boldsymbol{d} \rangle=\sum_{j}d_{j}\vert j \rangle , 
\end{equation} 
is a configuration correction
and $\hat{\mathcal{P}} = 1 - \vert 0\rangle\langle 0 \vert$ is the projection operator onto the complement to the reference state $\vert 0\rangle$.
The $\hat{\kappa}$ operator in Eq.~\eqref{srdftwavefunc} is the usual antisymmetric real singlet orbital-rotation operator
\begin{equation}
 \hat{\kappa} = \sum_{pq}\kappa_{pq}\hat{E}_{pq} = \sum_{p>q}\kappa_{pq}\left(\hat{E}_{pq} - \hat{E}_{qp}\right) \equiv \sum_{p>q}\kappa_{pq}\hat{E}^{-}_{pq}. \label{orbitalrotation}
\end{equation}
The charge-density- and spin-density-dependent terms in Eq.~\eqref{E_MCsrDFT} can now be expressed in a second-quantization formulation\cite{saue2002,salek2002}
in terms of their associated density operators 
\begin{align}
 \hat{\rho}_{X} (\mathbf{r}) & = \sum_{pq}\Omega_{pq}(\mathbf{r})\hat{O}^{X}_{pq}  ,
\label{density_operator_DFT}
\end{align}
with $\Omega_{pq}(\mathbf{r}) = \phi^{*}_{p}(\mathbf{r}) \phi_{q}(\mathbf{r})$.
We have in Eq.~\eqref{density_operator_DFT} introduced a nomenclature that will turn convenient in the next sections
for equations which are otherwise identical for charge-density and spin-density operators.
In this nomenclature, $X=C$ is used for the regular charge-density operator ($\hat{O}^{C}_{pq}\equiv\hat{E}_{pq}$), while $X = S$ denotes 
the spin-density operator
\begin{equation}
  \hat{O}^{S}_{pq} \equiv \hat{T}_{pq}  = \hat{a}^{\dagger}_{p\alpha}\hat{a}_{q\alpha} - \hat{a}^{\dagger}_{p\beta}\hat{a}_{q\beta} ,  \label{T_10} 
\end{equation} 
which is required to describe spin-density effects (in a non-relativistic framework).
We note that for spin-restricted models as considered here, the charge-density operator is a singlet operator
and the spin-density operator is a triplet operator of $M_S=0$ type. 
The wave function can be of any spin symmetry $S$, however, as in Kohn-Sham DFT, in practice the wave function should correspond to a spin component $M_S=S$ or $M_S=-S$
to give an appropriate spin density for the approximate spin-polarized exchange-correlation functional.
The electron charge density $\rho_{C}(\mathbf{r},\bm{\lambda})$ and electron spin density  
$\rho_S(\mathbf{r},\bm{\lambda})$
are obtained as expectation values according to
\begin{align}
\rho_X (\mathbf{r},\bm{\lambda}) = \langle \Psi^{\text{lr}}(\bm{\lambda}) \vert \hat{\rho}_{X}(\mathbf{r})\vert \Psi^{\text{lr}}(\bm{\lambda}) \rangle 
=  \sum_{pq}\Omega_{pq}(\mathbf{r}) D^{X}_{pq}(\bm{\lambda}) , \label{density_operator_MCsrDFT}
\end{align}
where $D^{X}_{pq}(\bm{\lambda})$ is the $(p,q)$ element of the one-electron reduced (spin-)density matrix
\begin{equation}
 D^{X}_{pq}(\bm{\lambda}) =  \langle\Psi^{\text{lr}}(\bm{\lambda}) \vert \hat{O}^{X}_{pq}\vert \Psi^{\text{lr}}(\bm{\lambda}) \rangle  . 
\end{equation}
The $E_{\text{sr-H}}[\rho_C]$ and $E_{\text{sr-xc}}[\rho_C,\rho_S]$ terms in Eq.~\eqref{E_MCsrDFT} can now be defined. 
The former is the short-range Hartree energy which depends only on the total density matrix
\begin{equation}
 E_{\text{sr-H}}[\rho_{C}] = \frac{1}{2} \sum_{pq,rs} D^{C}_{pq}(\bm{\lambda})\, g^{\text{sr}}_{pqrs}\, D^{C}_{rs}(\bm{\lambda}) 
 \equiv \frac{1}{2} \sum_{pq} D^{C}_{pq}(\bm{\lambda})\,j^{\text{sr}}_{pq}(\bm{\lambda})  , \label{E_H_energy_QC_derivation}
\end{equation}
where
$g^{\text{sr}}_{pqrs}$ are 
the short-range two-electron integrals (defined with the short-range interaction in Eq.~\eqref{lrsrpart}).  
The final term, $E_{\text{sr-xc}}[\rho_{C},\rho_S]$, in Eq.~\eqref{E_MCsrDFT} is the short-range exchange-correlation functional. This term 
has an explicit dependence on both the charge and spin densities
\begin{equation}
 E_{\text{sr-xc}}[\rho_{C}, \rho_S ] =  \int e_{\text{sr-xc}}(\rho_{C}(\mathbf{r},\bm{\lambda}),\rho_S(\mathbf{r},\bm{\lambda}) )\text{d}\mathbf{r} ,  
\label{E_xc}
\end{equation}
in accordance with Ref.~\citenum{saue2002}, where $e_{\text{sr-xc}}(\rho_{C}(\mathbf{r},\bm{\lambda}),\rho_S(\mathbf{r},\bm{\lambda}) )$
is the short-range exchange-correlation energy density. 
It should be noted that the expression in Eq.~\eqref{E_xc}
assumes the 
local-density approximation of the short-range exchange-correlation functional (srLDA)\cite{paziani2006}.
For brevity we only consider srLDA here, but 
we have also implemented the code for functionals based on a generalized-gradient approximation (GGA). The spin-dependent GGAs additionally depend on the electron charge density gradient and the electron spin density gradient.
The additional terms are structurally similar to the srLDA terms and will be given in Appendix \ref{app_GGA}. 

The MC--srDFT wave function is optimized with the restricted-step second-order MCSCF optimization 
algorithm\cite{jensen1984a,jensen1984b,jensen1986,jensen1987,jensen1994} as implemented in {\sc DALTON}.\cite{DALTON2016,WCMS:WCMS1172}
This algorithm is based on a second-order Taylor expansion 
of the electronic energy in the wave-function parameters, $\bm{\lambda}$, around $\bm{\lambda}=\bm{0}$
\begin{align}
 E(\bm{\lambda}) = E_0 + \bm{g}^{T}\bm{\lambda} + \frac{1}{2}\bm{\lambda}^{T}\bm{H}\bm{\lambda} + \cdots  .  \label{energy_taylor}
\end{align}
The electronic gradient, $\bm{g}$, and electronic Hessian, $\bm{H}$, are blocked according to the
configurational and orbital parameters of the wave function. Thus the gradient reads  
\begin{align}
 \bm{g} & =  \left(\begin{array}{c} \bm{g}^{c} \\ \bm{g}^{o} \end{array} \right) , \label{srDFT_gradient} 
\end{align}
and each gradient element (configurational or orbital) has both WFT (lr) and DFT (sr) contributions
\begin{align}
g_{i} & = g_{\text{lr},i} + g_{\text{sr-H},i} + g_{\text{sr-xc},i} \notag \\
& = \frac{\partial E_{\text{lr}}  }{\partial \lambda_i }  +   \frac{\partial E_{\text{sr-H}}[\rho_{C}] }{\partial \lambda_i }  
+   \frac{\partial E_{\text{sr-xc}}[\rho_{C},\rho_S] }{\partial \lambda_i }  .\label{g_elec_total} 
\end{align}
The electronic Hessian in Eq.~\eqref{energy_taylor} is evaluated as\cite{jensen1994} 
 $\bm{H} = \bm{P}\bm{K}\bm{P}$, 
where $\bm{P}$ denotes the matrix representation of the $\hat{\mathcal{P}}$ operator in Eq.~\eqref{srdftwavefunc}
and $\bm{K}$ has WFT and DFT contributions 
\begin{align}
 K_{ij}  & = K_{\text{lr},ij} + K_{\text{sr-H},ij} + K_{\text{sr-xc},ij}  \notag \\
& = \frac{\partial^{2}E_{\text{lr}}  }{\partial \lambda_i \partial\lambda_j}  +   \frac{\partial^{2}E_{\text{sr-H}}[\rho_{C}] }{\partial \lambda_i \partial\lambda_j}  
+   \frac{\partial^{2}E_{\text{sr-xc}}[\rho_{C},\rho_S] }{\partial \lambda_i \partial\lambda_j} .  \label{K_elec_total} 
\end{align}
In the actual implementation the Hessian matrix is never constructed explicitly. Instead, Hessian contributions to the wave-function optimization process
are obtained in a direct fashion based on trial vectors
\begin{align}
 \bm{\sigma}_{n} &= \bm{PKP}\,\bm{b}_{n} =  \bm{PK}\,\bm{b}_{n} = \bm{P}
 \left(\begin{array}{cc} \bm{K}^{cc} & \bm{K}^{co} \\
                          \bm{K}^{oc} & \bm{K}^{oo} \end{array} \right)
\left(\begin{array}{cc} \bm{b}^{c}_{n} \\ \bm{b}^{o}_{n} \end{array}\right)  \label{srDFT_Hessian},   
\end{align}
where the individual contributions to $\bm{\sigma}_{n}$ can be written in terms of modified Fock matrices (see Section \ref{implement})
and the second equal sign stems from that we require each trial vector to fulfill $\bm{P}\,\bm{b}_{n} =  \bm{b}_{n}$.
The explicit expressions for the new short-range DFT contributions to the individual gradient and $\bm{\sigma}_{n}$ element types are derived below. 
The MCSCF long-range contributions are equivalent to the terms in the regular MCSCF method which together with  the second-order optimization algorithm are documented in the original works describing
the {\sc Dalton} implementation.\cite{jensen1984a,jensen1984b,jensen1986,jensen1987,jensen1994}

\subsection{The sr-DFT contributions to the electronic gradient}

The short-range exchange-correlation contributions to the gradient in Eq.~\eqref{g_elec_total} can be obtained as    
\begin{equation}
g_{\text{sr-xc},i} = \frac{\partial E_{\text{sr-xc}}[\rho_{C},\rho_S]}{{\partial \lambda_i}} 
= \int \frac{\partial e_{\text{sr-xc}}(\rho_{C}(\mathbf{r}),\rho_S(\mathbf{r}))}{\partial\lambda_i } \text{d}\mathbf{r} .  \label{e_xc_chain_rule_1}
\end{equation}
The derivative within the kernel in Eq.~\eqref{e_xc_chain_rule_1} is obtained through the chain rule  
\begin{align}
 \frac{\partial e_{\text{sr-xc}}(\rho_{C}(\mathbf{r}),\rho_S(\mathbf{r}))}{\partial\lambda_i } 
&= \frac{\partial e_{\text{sr-xc}}(\rho_{C}(\mathbf{r}),\rho_S(\mathbf{r}))}{\partial\rho_{C}} \frac{\partial \rho_{C}(\mathbf{r})}{\partial \lambda_i} + 
 \frac{\partial e_{\text{sr-xc}}(\rho_{C}(\mathbf{r}),\rho_S(\mathbf{r}))}{\partial \rho_{S}} \frac{\partial \rho_{S}(\mathbf{r})}{\partial \lambda_i} \notag \\
&= 
\sum_{pq} \Biggl[ \left(\frac{\partial e_{\text{sr-xc}}(\rho_{C}(\mathbf{r}),\rho_S(\mathbf{r}))}{\partial\rho_{C}} \Omega_{pq}(\mathbf{r})\right) \frac{\partial D^{C}_{pq}(\bm{\lambda})}{\partial  \lambda_i}  
\notag\\
&\;\;\;\;\;\;\;\;\;\;\;\;
+ \left(\frac{\partial e_{\text{sr-xc}}(\rho_{C}(\mathbf{r}),\rho_S(\mathbf{r}))}{\partial\rho_S} \Omega_{pq}(\mathbf{r})\right) \frac{\partial D^S_{pq}(\bm{\lambda})}{\partial  \lambda_i} \Biggl],
\label{e_xc_chain_rule_3}
\end{align}
where $\rho_{C}(\mathbf{r})$ and $\rho_S (\mathbf{r})$ from Eq.~(\ref{density_operator_MCsrDFT}) have 
been inserted. 
As usual in MCSCF schemes, the orbital and configurational parts of the gradient leads to different computational expressions.
The orbital part of the short-range exchange-correlation gradient becomes 
\begin{align}
\frac{\partial E_{\text{sr-xc}}[\rho_{C},\rho_S] }{\partial \kappa_{rs} } =  g^{C}_{\text{sr-xc},rs} + g^S_{\text{sr-xc},rs}, 
\label{gradient_orb}  
\end{align}
where the two gradient terms are
\begin{subequations}
\begin{equation}
 g^{C}_{\text{sr-xc},rs}    = \langle 0\vert\, [\hat{E}^{-}_{rs},\hat{V}^{C,g}_{\text{sr-xc}}]\, \vert0\rangle = 
    2\langle 0\vert\, [\hat{E}_{rs},\hat{V}^{C,g}_{\text{sr-xc}}]\, \vert0\rangle,
\label{gradient_orb_1}  
\end{equation}
\begin{equation}
 g^S_{\text{sr-xc},rs}   = \langle 0\vert\, [\hat{E}^{-}_{rs},\hat{V}^{S,g}_{\text{sr-xc}}]\, \vert0\rangle  = 
    2\langle 0\vert\, [\hat{E}_{rs},\hat{V}^{S,g}_{\text{sr-xc}}]\, \vert0\rangle, 
\label{gradient_orb_2}  
\end{equation}
\label{gradient_orb_12}  
\end{subequations}
both defined in terms of an effective operator ($X=C$ and $X=S$, respectively)
\begin{align}
 \hat{V}^{X,g}_{\text{sr-xc}} = \sum_{pq} \left( \int\frac{\partial e_{\text{sr-xc}}(\rho_{C}(\mathbf{r}),\rho_{S}(\mathbf{r}))}{\partial \rho_{X}}\Omega_{pq}(\mathbf{r})\text{d}\mathbf{r}\right)\hat{O}^{X}_{pq} \equiv \sum_{pq}V^{X,g}_{\text{sr-xc},pq}\hat{O}^{X}_{pq} .
\label{Vg_eff}
\end{align}
We note that the $g^{C}_{\text{sr-xc},rs}$ term in Eq.~\eqref{gradient_orb_1} is the term already present in the closed-shell formalism, although in that case the functional kernel in the effective operator does not 
depend on $\rho_S$. The $g^{S}_{\text{sr-xc},rs}$ term in
Eq.~\eqref{gradient_orb_2} exclusively occurs in the open-shell formalism. The two gradient terms are commonly denoted $\textbf{g}^X_{\text{sr-xc}}$.  
The configurational part of the gradient can be defined similarly
\begin{align}
 \frac{\partial E_{\text{sr-xc}}[\rho_{C},\rho_S] }{\partial d_{j} } = g^{C}_{\text{sr-xc},j} +  g^{S}_{\text{sr-xc},j} , 
 \label{gradient_xc_conf_1}
\end{align}
where
\begin{align}
 g^{X}_{\text{sr-xc},j}  
   &= 2\left( \langle j \vert \hat{V}^{X,g}_{\text{sr-xc}}\vert 0 \rangle 
- c_j \langle 0 \vert \hat{V}^{X,g}_{\text{sr-xc}}\vert 0\rangle \right),   \label{gradient_xc_conf_2}
\end{align}
and we again have employed the effective operators defined in Eq.~\eqref{Vg_eff}.

Finally, we also briefly describe the gradient contribution from the short-range Hartree term,  $E_{\text{sr-H}}[\rho_{C}]$, in  Eq.~\eqref{E_H_energy_QC_derivation}  
\begin{align}
g_{\text{sr-H},i} & = \frac{\partial E_{\text{sr-H}}[\rho_{C}]}{\partial\lambda_i} 
 = \sum_{pq}  j^{\rm sr}_{pq}(\bm{\lambda})  \frac{\partial D^{C}_{pq}(\bm{\lambda})}{\partial\lambda_i}  . \label{gradient_EH_1}
\end{align}
The configurational and orbital gradient contributions are identical to Eqs.~\eqref{gradient_orb_1} and \eqref{gradient_xc_conf_2}, except that the effective operator 
\begin{align}
\hat{V}^{g}_{\text{sr-H}} =  \sum_{pq}j_{pq}^\text{sr} \hat{E}_{pq} , 
\end{align}
replaces $\hat{V}^{C,g}_{\text{sr-xc}}$. Combined, we obtain the effective operators for the total short-range Hartree-exchange-correlation functional as 
\begin{align}
\hat{V}^{X,g}_{\text{sr-Hxc}} = \left\{\begin{array}{lcc} \hat{V}^{g}_{\text{sr-H}} + \hat{V}^{C,g}_{\text{sr-xc}}, & & X=C   \\
          \hat{V}^{S,g}_{\text{sr-xc}},         & &  X=S 
          \end{array} \right. \label{V_g_total}
\end{align}
The total  gradient vector (from the short-range DFT part) can thus be obtained from 
\begin{align}
\mathbf{g}_{\text{sr-Hxc}} = \mathbf{g}^{C}_{\text{sr-Hxc}} + \mathbf{g}^{S}_{\text{sr-xc}} . 
\label{gradient_tot}
\end{align}

\subsection{The sr-DFT contributions to the electronic Hessian sigma vectors}

As for the gradient in the previous subsection, we focus on the exchange-correlation term 
since the second derivative of $E_{\text{lr}}$ in Eq.~(\ref{E_MCsrDFT}) is equivalent to the terms from regular MCSCF\cite{jensen1984a,jensen1984b,jensen1986,jensen1987}, and the 
Hartree term can be obtained from the total density matrix and has no explicit spin-dependent terms.    
The Hessian contributions from the exchange-correlation functional are given by
\begin{align}
K_{\text{sr-xc},ij} =
\frac{\partial^2 E_{\text{sr-xc}}[\rho_{C},\rho_S ]}{ \partial \lambda_i \partial \lambda_j } 
& = \int
\frac{\partial^2 e_{\text{sr-xc}}(\rho_{C}(\mathbf{r}),\rho_S(\mathbf{r}) )}{\partial \lambda_i \partial \lambda_j } \text{d}\mathbf{r} , 
\label{E_xc_Hes}
\end{align}	
and from the chain rule
\begin{align}
\frac{\partial^2 e_{\text{sr-xc}}(\rho_{C}(\mathbf{r}),\rho_{S}(\mathbf{r}))}{\partial \lambda_i \partial \lambda_j } & = 
\frac{\partial^2 e_{\text{sr-xc}}(\rho_{C}(\mathbf{r}),\rho_S(\mathbf{r}))}{ \partial \rho_{C}^2}
\frac{\partial \rho_{C}(\mathbf{r})}{\partial  \lambda_i }
\frac{\partial \rho_{C}(\mathbf{r})}{\partial \lambda_j } +  
\frac{\partial^2 e_{\text{sr-xc}}(\rho_{C}(\mathbf{r}),\rho_{S}(\mathbf{r}))}{ \partial \rho_{C}  \partial \rho_{S}}
\frac{\partial \rho_{S}(\mathbf{r})}{\partial  \lambda_i }
\frac{\partial \rho_{C}(\mathbf{r})}{\partial \lambda_j } 
 \notag \\ 
& +  
\frac{\partial^2 e_{\text{sr-xc}}(\rho_{C}(\mathbf{r}),\rho_{S}(\mathbf{r}))}{ \partial \rho_{S}  \partial \rho_{C} }
\frac{\partial \rho_{C}(\mathbf{r})}{\partial  \lambda_i }
\frac{\partial \rho_{S}(\mathbf{r})}{\partial \lambda_j } 
 + \frac{\partial^2 e_{\text{sr-xc}}(\rho_{C}(\mathbf{r}),\rho_{S}(\mathbf{r}))}{ \partial \rho_{S}^2}
\frac{\partial \rho_{S}(\mathbf{r})}{\partial  \lambda_i }
\frac{\partial \rho_{S}(\mathbf{r})}{\partial \lambda_j }
\notag \\
& +  \frac{\partial e_{\text{xc}}\rho_{C}(\mathbf{r}),\rho_{S}(\mathbf{r}))}{\partial \rho_{C}}\frac{\partial^2 \rho_{C}(\mathbf{r})}{ \partial \lambda_i \partial \lambda_j} + 
 \frac{\partial e_{\text{sr-xc}}(\rho_{C}(\mathbf{r}),\rho_{S}(\mathbf{r}))}{\partial \rho_{S}}\frac{\partial^2 \rho_{S}(\mathbf{r})}{ \partial \lambda_i \partial \lambda_j} . 
\label{exc_hessian_chain_rule_2}
\end{align}
In practice we utilize a direct Hessian technique [cf.~Eq.~(\ref{srDFT_Hessian})] where the quantity in Eq.~\eqref{E_xc_Hes}  
is contracted with configurational ($b^c_{j}$) or orbital ($b^o_{pq}$) trial vectors.
The direct Hessian contributions from the $E_{\text{sr-xc}}[\rho_{C},\rho_S ]$ term thus consist of four different types  
\begin{subequations}
\begin{align}
 \sum_{j}\frac{\partial^2 E_{\text{sr-xc}}[\rho_{C},\rho_S ]}{\partial d_i \partial d_j}\ b^{c}_j & = \sigma^{C;c}_{\text{sr-xc},i} + \sigma^{S;c}_{\text{sr-xc},i} \label{E_xc_HesCI_c_c} \\ 
 \sum_{j}\frac{\partial^2 E_{\text{sr-xc}}[\rho_{C},\rho_S ]}{\partial \kappa_{pq} \partial d_j}\ b^{c}_{j}   &= \sigma^{C;c}_{\text{sr-xc},pq} +  \sigma^{S;c}_{\text{sr-xc},pq} \label{E_xc_HesCI_c_orb}  \\
 \sum_{r>s}\frac{\partial^2 E_{\text{sr-xc}}[\rho_{C},\rho_S ]}{\partial d_i \partial \kappa_{rs} }\ b^{o}_{rs} &= \sigma^{C;o}_{\text{sr-xc},i} + \sigma^{S;o}_{\text{sr-xc},i} \label{E_xc_HesCI_o_c}  \\
 \sum_{r>s}\frac{\partial^2 E_{\text{sr-xc}}[\rho_{C},\rho_S ]}{\partial \kappa_{pq} \partial \kappa_{rs}}\ b^{o}_{rs}  & = \sigma^{C;o}_{\text{sr-xc},pq} +  \sigma^{S;o}_{\text{sr-xc},pq}. \label{E_xc_HesCI_o_orb} 
\end{align}  
\end{subequations}
The expressions in Eqs.~\eqref{E_xc_HesCI_c_c}--\eqref{E_xc_HesCI_o_orb} are  obtained by inserting $\rho_{C}(\mathbf{r})$ and $\rho_{S}(\mathbf{r})$ from Eq.~\eqref{density_operator_MCsrDFT} into Eq.~\eqref{exc_hessian_chain_rule_2}, and 
inserting the resulting expression into Eq.~\eqref{E_xc_Hes} and contracting with configurational or orbital 
trial vectors, respectively. We sketch the two steps below; the insertion of the densities gives    
\begin{align}
\frac{\partial^2 E_{\text{sr-xc}}[\rho_{C},\rho_S]}{ \partial \lambda_i \partial \lambda_j } 
& = 
\sum_{X,Y=C,S}\sum_{pq,rs}  \frac{\partial D^{Y}_{rs}(\bm{\lambda})}{\partial  \lambda_j } 
\left( \int  \frac{\partial^2 e_{\text{sr-xc}}(\rho_{C}(\mathbf{r}),\rho_S(\mathbf{r}))}{ \partial\rho_{Y}\partial\rho_{X}}\Omega_{pq}(\mathbf{r}) 
\Omega_{rs}(\mathbf{r}) \text{d}\mathbf{r} \right)
 \frac{\partial D^{X}_{pq}(\bm{\lambda})}{\partial \lambda_i } \notag \\
& + \sum_{X=C,S}\sum_{pq}\left(\int \frac{\partial e_{\text{sr-xc}}(\rho_{C}(\mathbf{r}),\rho_S(\mathbf{r}))}{\partial \rho_{X}}\Omega_{pq}(\mathbf{r})  \text{d}\mathbf{r} \right)
\frac{\partial^2 D^{X}_{pq}(\bm{\lambda})}{ \partial \lambda_i \partial \lambda_j} .
\label{E_xc_Hes_2}
\end{align}
By contraction of the above equation with the trial vectors, the configurational part becomes 
\begin{align}
 \sigma^{X;c}_{\text{sr-xc},i} =&  \sum_{Y=C,S} \sum_{pq,rs}   \left\{ \int 
  \frac{\partial^2 e_{\text{sr-xc}}(\rho_{C}(\mathbf{r}),\rho_S(\mathbf{r}))}{ \partial \rho_{Y}\partial\rho_{X}}  \sum_{j}\left(\frac{\partial D^{Y}_{rs}(\boldsymbol{\lambda})}{\partial d_{j}}b^{c}_{j}\ \Omega_{rs}(\mathbf{r}) \right) \Omega_{pq}(\mathbf{r}) \text{d}\mathbf{r} \right\} \frac{\partial D^{X}_{pq}(\bm{\lambda})}{\partial \lambda_{i}}   \notag\\
+& \sum_{pq}   
\left\{\int \frac{\partial e_{\text{sr-xc}}(\rho_{C}(\mathbf{r}),\rho_S(\mathbf{r}))}{\partial \rho_{X}}\Omega_{pq}(\mathbf{r})  \text{d}\mathbf{r} \right\} 
\sum_{j}\frac{\partial^{2} D^{X}_{pq}(\boldsymbol{\lambda})}
{\partial \lambda_{i} \partial d_{j}}  b^{c}_j    ,
\label{E_xc_HesCI_c} 
\end{align}
while the orbital part becomes 
\begin{align}
 \sigma^{X;o}_{\text{sr-xc},i} = & \sum_{Y=C,S} \sum_{pq,rs}  
\left\{ \int \frac{\partial^2 e_{\text{sr-xc}}(\rho_{C}(\mathbf{r}),\rho_S(\mathbf{r}))}{\partial\rho_{Y}\partial\rho_{X}}  \sum_{t>u}\left(\frac{\partial D^{Y}_{rs}(\boldsymbol{\lambda})}{\partial \kappa_{tu}}b^{o}_{tu}\ \Omega_{rs}(\mathbf{r}) \right)  \Omega_{pq}(\mathbf{r})\text{d}\mathbf{r} \right\}
\frac{\partial D^{X}_{pq}(\boldsymbol{\lambda})}{\partial \lambda_{i}} \notag\\
+& \sum_{pq}
\left\{\int \frac{\partial e_{\text{sr-xc}}(\rho_{C}(\mathbf{r}),\rho_S(\mathbf{r}))}{\partial \rho_{X}} \Omega_{pq}(\mathbf{r})  \text{d}\mathbf{r} \right\} \sum_{t>u} \frac{\partial^{2} D^{X}_{pq}(\bm{\lambda})}
{\partial \lambda_{i} \partial \kappa_{tu}}   b_{tu}^o.   \label{E_xc_Hes_orbital} 
\end{align} 
The first term in both Eqs.~\eqref{E_xc_HesCI_c} and \eqref{E_xc_Hes_orbital} occur due to the non-linearity of the srDFT energy functional. 
These terms 
can be formulated as transformed gradient terms with effective operators as summarized below. 
The last term in both Eqs.~\eqref{E_xc_HesCI_c} and \eqref{E_xc_Hes_orbital} are similar to the one-electron Hessian part for a regular MCSCF calculation,
the only difference being that they employ 
different integrals (over the first-order derivative of the srDFT functional).
To define the effective operators employed in the first term of the two equations, we define the 
transition density matrices for configurational trial vectors 
  \begin{align}
D^{Y(1c)}_{rs} \equiv 
\sum_{j}\frac{\partial D^{Y}_{rs}(\boldsymbol{\lambda})}{\partial d_{j}}b^{c}_{j} & 
 = \sum_{j}\left(b^{c}_{j}\langle j \vert \hat{O}^{Y}_{rs}\vert 0 \rangle + \langle 0 \vert \hat{O}^{Y}_{rs}\vert j \rangle b^{c}_{j}\right) \notag \\
 & = \langle B \vert \hat{O}^{Y}_{rs}\vert 0\rangle + \langle 0 \vert \hat{O}^{Y}_{rs}\vert B\rangle , \label{CI_transformed_gradient}
\end{align}
where $\vert B\rangle$ is the state vector generated by the current trial vector: $\vert B\rangle = \sum_{j}b_j^c\vert j \rangle$. 
For the orbital trial vectors, we define the one-index transformed density matrix
\begin{align}
D^{Y(1o)}_{rs} \equiv 
\sum_{t>u}\frac{\partial D^{Y}_{rs}(\boldsymbol{\lambda})}{\partial \kappa_{tu}}   b^{o}_{tu}
 = \sum_{t}\left(D^Y_{ts}b^{o}_{rt} + D^Y_{rt}b^{o}_{st} \right).  \label{orb_transformed_gradient} 
\end{align}
The transformed configurational and orbital (spin-)density matrices, $\mathbf{D}^{Y(1c)}$ 
and $\mathbf{D}^{Y(1o)}$, are collectively denoted by $\mathbf{D}^{Y(1\lambda)}$. We can now define the integrals that comprise the effective operators in terms of  
the linearly transformed densities 
\begin{align}
 V^{X[1\lambda]}	_{\text{sr-xc},pq}[\mathbf{D}^{Y(1\lambda)}] =  \int \frac{\partial^2 e_{\text{sr-xc}}(\rho_{C}(\mathbf{r}),\rho_{S}(\mathbf{r}))}{ \partial \rho_{Y}
\partial \rho_{X}}\left(\sum_{rs} D^{Y(1\lambda)}_{rs}\Omega_{rs}(\mathbf{r}) \right) 
\Omega_{pq}(\mathbf{r}) \text{d}\mathbf{r}  , 
\end{align}
which is a generic term for the integrals in the curvy brackets of the first term in Eqs.~\eqref{E_xc_HesCI_c} and \eqref{E_xc_Hes_orbital}.  
With this generic definition of the integrals, we can define the two effective operators of either singlet or triplet types 
\begin{subequations}
\begin{align}
  \hat{V}^{C[1\lambda]}_{\text{sr-xc}} & = \sum_{pq}\Bigl( V^{C[1\lambda]}_{\text{sr-xc},pq}[\mathbf{D}^{C(1\lambda)}] 
+ V^{C[1\lambda]}_{\text{sr-xc},pq}[\mathbf{D}^{S(1\lambda)}] \Bigr) \hat{E}_{pq} \label{V_eff_1_xc_c} \\
  \hat{V}^{S[1\lambda]}_{\text{sr-xc}} & = \sum_{pq}\Bigl( V^{S[1\lambda]}_{\text{sr-xc},pq}[\mathbf{D}^{C(1\lambda)}] 
+ V^{S[1\lambda]}_{\text{sr-xc},pq}[\mathbf{D}^{S(1\lambda)}]\Bigr) \hat{T}_{pq}. 
\label{V_eff_1_xc_s}  
\end{align}	 
\end{subequations}
The direct Hessian terms in Eqs.~\eqref{E_xc_HesCI_c_c}--\eqref{E_xc_HesCI_o_orb} can now be written in the generic form  
\begin{subequations}
\begin{align}
 \sigma^{X;c}_{\text{sr-xc},i} & =   2\left( \langle i \vert   \hat{V}^{X[1c]}_{\text{sr-xc}} \vert 0 \rangle 
- c_i \langle 0 \vert   \hat{V}^{X[1c]}_{\text{sr-xc}} \vert 0\rangle \right) 
+ 2\left( \langle i \vert \hat{V}^{X,g}_{\text{sr-xc}} \vert B \rangle - \langle 0\vert \hat{V}^{X,g}_{\text{sr-xc}}  \vert 0 \rangle b^c_{j} \right)
         \label{E_xc_HesCI_c_c_2} \\
 \sigma^{X;c}_{\text{sr-xc},pq} & =  2  \langle 0\vert\, [\hat{E}_{pq}, \hat{V}^{X[1c]}_{\text{sr-xc}}]\, \vert0\rangle + 
2\langle 0 \vert [\hat{E}_{pq}^-,\hat{V}^{X,g}_{\text{sr-xc}}]\vert B \rangle
         \label{E_xc_HesCI_c_orb_2}   \\
 \sigma^{X;o}_{\text{sr-xc},i} & =   2  \left( \langle i \vert \hat{V}^{X[1o]}_{\text{sr-xc}} \vert 0 \rangle 
- c_i \langle 0 \vert  \hat{V}^{X[1o]}_{\text{sr-xc}} \vert 0\rangle \right)
+ 2 \langle i \vert \hat{\tilde{V}}^{X,g}_{\text{sr-xc}}\vert 0 \rangle
         \label{E_xc_HesCI_o_c_2} \\
 \sigma^{X;o}_{\text{sr-xc},pq} & =  \tilde{g}^{X}_{\text{sr-xc},pq} + \langle 0 \vert [\hat{E}_{pq}^-,\hat{V}^{X[1o]}_{\text{sr-xc}} + \hat{\tilde{V}}^{X,g}_{\text{sr-xc}}]\vert 0 \rangle + 
\frac{1}{2}\sum_{t}\Bigl( g^X_{\text{sr-xc},tp}b^o_{qt} - g^X_{\text{sr-xc},tq}b^o_{pt} \Bigr). 
         \label{E_xc_HesCI_o_orb_2} 
\end{align}
\end{subequations}
The orbital gradient elements, $g^X_{\text{sr-xc},pq}$, were defined in Eqs.~\eqref{gradient_orb_1}-\eqref{gradient_orb_2}, the $\hat{V}^{X,g}_{\text{sr-xc}}$ operators were 
given in Eq.~\eqref{Vg_eff}, and $\hat{\tilde{V}}^{X,g}_{\text{sr-xc}}$ is the one-index transformed form of the $\hat{V}^{X,g}_{\text{sr-xc}}$ operator in Eq.~\eqref{Vg_eff}. The one-index transformed operators employ transformed integrals\cite{jensen1986,jensen1994}, e.g., $\hat{\tilde{V}}^{X,g}_{\text{sr-xc}}$ is equivalent to Eq.~\eqref{Vg_eff}, but with the $V^{X,g}_{\text{sr-xc},pq}$ integrals replaced by 
\begin{align}
\tilde{V}^{X,g}_{\text{sr-xc},pq} = \sum_{t}\bigl(V^{X,g}_{\text{sr-xc},tq}b^o_{pt} + V^{X,g}_{\text{sr-xc},pt} b^{o}_{qt} \bigr).
\end{align}
Similarly, the expression for the transformed gradient elements, $\tilde{g}^X_{\text{sr-xc},pq}$, is equivalent to Eqs.~\eqref{gradient_orb_1}--\eqref{gradient_orb_2}, but with $\hat{V}^{X,g}_{\text{sr-xc}}$ replaced with $\hat{\tilde{V}}^{X,g}_{\text{sr-xc}}$. 
The terms with $\hat{V}^{X,g}_{\text{sr-xc}}$ and $\hat{\tilde{V}}^{X,g}_{\text{sr-xc}}$ operators arise from the last terms of Eqs.~\eqref{E_xc_HesCI_c} and \eqref{E_xc_Hes_orbital}, cf.~the part within curly brackets. For these terms, the regular Hessian structure can be discerned. 

At this point, it is noted that the $\bm{\sigma}^{S;c}$ and  $\bm{\sigma}^{S;o}$ terms in Eqs.~\eqref{E_xc_HesCI_c_c_2}--\eqref{E_xc_HesCI_o_orb_2} 
enter exclusively for open-shell systems, and are thus part of the extension for this work.
Accordingly, with respect to the Hessian this work concerns
the implementation of the effective operator of triplet type, i.e.~Eq.~\eqref{V_eff_1_xc_s}, employed to define the  $\bm{\sigma}^{S;c}$ and  $\bm{\sigma}^{S;o}$ terms. In addition, 
the term that depends on $\mathbf{D}^{S(1\lambda)}$ in Eq.~\eqref{V_eff_1_xc_c} also enters exclusively for open-shell systems and has been implemented here.   

In addition to the contributions from the exchange-correlation functional, the direct Hessian terms also contain contributions from the short-range Hartree term, $E_{\text{sr-H}}[\rho_{C}]$, in Eq.~\eqref{E_H_energy_QC_derivation} 
\begin{align}
 \frac{\partial^{2} E_{\text{sr-H}}[\rho_{C}]}{\partial \lambda_j \partial \lambda_{i}}  
  = &  \sum_{pq,rs} \frac{\partial D^{C}_{rs}(\boldsymbol{\lambda})}{\partial \lambda_{j}} 
 g^{\text{sr}}_{pqrs} \frac{\partial D^{C}_{pq}(\bm{\lambda})}{\partial \lambda_i }
+  
\sum_{pq,rs} D^{C}_{rs}(\bm{\lambda})   g^{\text{sr}}_{pqrs}  \frac{\partial^{2} D^{C}_{pq}(\boldsymbol{\lambda})}
{\partial \lambda_{j}\partial \lambda_{i}}   .
\label{E_H_Hes} 
\end{align}
As for the short-range exchange-correlation contributions,
the direct Hessian short-range Hartree contributions can be obtained by contraction of 
configurational and orbital trial vectors, respectively. The terms have the same structure as Eqs.~\eqref{E_xc_HesCI_c_c_2}--\eqref{E_xc_HesCI_o_orb_2}, but without any spin-dependence,
 and can be defined employing the effective operator 
\begin{align}
 \hat{V}^{C[1\lambda],c}_{\text{sr-H}} = \sum_{pq}j^{C(1\lambda)}_{\text{sr},pq}[\mathbf{D}^{C(1\lambda)}]\hat{E}_{pq} \label{E_H_eff_operator} 
&& \text{with} 
&& j^{C(1\lambda)}_{\text{sr},pq}[\mathbf{D}^{C(1\lambda)}] = \sum_{rs}D^{C(1\lambda)}_{rs} g^{\text{sr}}_{rspq} . 
\end{align}
We can combine the operators utilized for the short-range Hartree and short-range exchange-correlation terms in a common, effective operator 
\begin{align}
 \hat{V}^{X[1\lambda]}_{\text{sr-Hxc}} =  \left\{\begin{array}{lcc} \hat{V}^{C[1\lambda]}_{\text{sr-H}} +  \hat{V}^{C[1\lambda]}_{\text{sr-xc}}, & & X=C   \\
\hat{V}^{S[1\lambda]}_{\text{sr-xc}},         & &  X=S 
          \end{array} \right. , 
\label{Veff_common}
\end{align}
which replaces $\hat{V}^{X[1\lambda]}_{\text{sr-xc}}$ in Eqs.~\eqref{E_xc_HesCI_c_c_2}--\eqref{E_xc_HesCI_o_c_2}. 

\section{Implementation}\label{implement}

The configurational and orbital gradients for MC--srDFT were given in Eqs.~\eqref{gradient_orb}--\eqref{gradient_tot}  
and depend on $\hat{V}^{C,g}_{\text{sr-xc}}$ and $\hat{V}^{S,g}_{\text{sr-xc}}$ operators [Eq.~(\ref{Vg_eff})]. 
The implementations of the equations for the gradient and (direct) Hessian in regular MCSCF within {\sc DALTON} has been 
described previously.\cite{jensen1984a,jensen1984b,jensen1986} A central part of this implementation builds on the generation of generalized Fock-type matrices 
\begin{align}
 f_{pq} = \sum_{r}D^C_{pr}\,h_{qr} + \sum_{rst} P_{pr,st}\,g_{qr,st} \label{general_fock} ,
\end{align}
where $\mathbf{P}$ is the reduced two-electron density matrix with elements $P_{pr,st} = \langle \Psi| \hat{e}_{pr,st} | \Psi \rangle$.
In the following, we will in addition to the general orbitals (with indices $p,q,r,s$) need to denote both inactive and active orbitals, for which we use the indices $i,j$ and $u, v, x, y$, respectively. 
In terms of the generalized Fock matrix, the orbital gradient becomes  
\begin{align}
 g^{o}_{rs} = 2(f_{rs} - f_{sr}). \label{orb_grad_fckmat}
\end{align}
The direct Hessian is constructed in terms of modified gradient expressions, which in Ref.~\citenum{jensen1986} are denoted transition and one-index transformed gradient expressions.  
These can again be constructed from Eq.~\eqref{general_fock} by employing a one-index transformed density matrix or a transition density matrix instead of the reference density matrix.\cite{siegbahn1984,jensen1986} 
The only non-zero long-range contributions to the Fock matrix elements are \cite{jensen1986}
\begin{align}
 f^{\text{lr}}_{iq} & = 2(f^{I,\text{lr}}_{iq} + f^{A,\text{lr}}_{iq}) \label{fd_mat} \\
 f^{\text{lr}}_{vq} & = \sum_{u}D^C_{uv}f^{I,\text{lr}}_{uq} + Q^{\text{lr}}_{vq} \label{fd_mat_2} , 
\end{align}
where $Q^{\text{lr}}_{vq} = \sum_{uxy} P_{vu,xy}\,g_{qu,xy}^\text{lr}$.
The Fock matrices $\mathbf{f}^{I,lr}$ and $\mathbf{f}^{A,lr}$ are called the \textit{inactive} and \textit{active} Fock matrix, respectively.
They are defined as in Eqs.~(3.11) and (3.12) of Ref.\citenum{jensen1986}, but with long-range two-electron integrals. For an MC--srDFT description, the generalized Fock matrix elements must be altered to accommodate the potential contributions from the short-range DFT functional.
Eqs.~\eqref{fd_mat} and \eqref{fd_mat_2} become
\begin{align}
 f_{iq} &=  f^{\text{lr}}_{iq} + V^{C,g}_{\text{sr-Hxc},iq} + V^{S,g}_{\text{sr-xc},iq} \label{fd_mat_3}   \\
 f_{vq} & = \sum_{u}D^c_{uv}\left(f^{I,\text{lr}}_{uq} + V^{C,g}_{\text{sr-Hxc},uq} + V^{S,g}_{\text{sr-xc},uq}\right)  +  Q^{\text{lr}}_{vq} , \label{fd_mat_4}
\end{align}
where $V^{X,g}_{\text{sr-Hxc},pq}$ are the matrix elements of the operator in Eq.~\eqref{V_g_total}. Depending on the employed (spin-)density matrix, Eqs.~\eqref{fd_mat_3} and \eqref{fd_mat_4} can be either a 
regular (spin-)Fock-matrix, a transition-(spin-)Fock matrix or an one-index transformed (spin-)Fock matrix.  

The auxiliary routines that construct Fock matrices were modified to accommodate short-range functionals in connection with previous work\cite{pedersenphd2004}. These auxiliary routines have been extended with the construction of the spin-density matrix required for the spin-dependent terms of $V^{X,g}_{\text{sr-Hxc},pq}$. The direct Hessian [Eqs. (\ref{E_xc_HesCI_c_c_2})--(\ref{E_xc_HesCI_o_orb_2})] is comprised of terms that include $\mathbf{D}^{X(1\lambda)}$  (linear transformed terms) and ``regular'' Hessian terms. The former terms required slight modification for this work, whereas the latter terms are automatically included after the above-mentioned modifications of the gradient routines. 

Finally, we have added code for the spin-dependent 
short-range exchange-correlation functionals from Ref.~\citenum{paziani2006} and the spin-dependent gradient-corrected functional of Ref.~\citenum{goll2006} (see next section). This includes code for the numerical evaluation of the gradient and Hessian of the short-range functional kernel with respect to the spin density.

\section{Computational Details} \label{comp}

The theory presented in the previous sections was implemented in a development version of the {\sc DALTON} 2016 program.\cite{DALTON2016,WCMS:WCMS1172} 
All CAS--srDFT calculations were carried out with this new implementation. We have employed CAS--srDFT for two molecules, namely \ce{O2} and \ce{[Fe(H2O)6]^{3+}}, exploiting point group symmetry: the $D_{2h}$ symmetry sub-group of $D_{\infty h}$ for \ce{O2}, and the $C_i$ symmetry group for \ce{[Fe(H2O)6]^{3+}}.

For \ce{O2} we focus on the $^3\Sigma^-_{g}$ and $^1 \Delta_{g}$ lowest-energy states.
The calculations on \ce{O2} were carried out at the experimental ground-state equilibrium bond distance ($R_{\text{eq}}=1.207$~\AA) and at
a stretched bond distance ($R_{\text{stretch}}=2.0$~\AA) with the cc-pVTZ basis set.\cite{bs890115d,bs701001d}   
For \ce{O2}, we have employed three CAS($n,m$) spaces with $n$ electrons in $m$ orbitals; CAS(8,6), CAS(12,8) and CAS(12,16). These spaces correspond to including the orbitals $3\sigma_g^2 1\pi^4_{u}1\pi^{*2}_{g}3\sigma_u^{*0}$, $2\sigma_g^{2}2\sigma_u^{*2} 3\sigma_g^2 1\pi^4_{u} 1\pi^{*2}_{g}3\sigma^{*0}$, and $2\sigma_g^{2}2\sigma_u^{*2} 3\sigma_g^2 1\pi^4_{u} 1\pi^{*2}_{g}3\sigma_u^{*0}4\sigma_g^0 4\sigma_u^{*0}5\sigma_g^0 2\pi^0_{u}2\pi^{*0}_{g}5\sigma_u^{*0}$, respectively, where the occupations are for the reference configuration. 

\begin{figure}
    \includegraphics[scale=0.20]{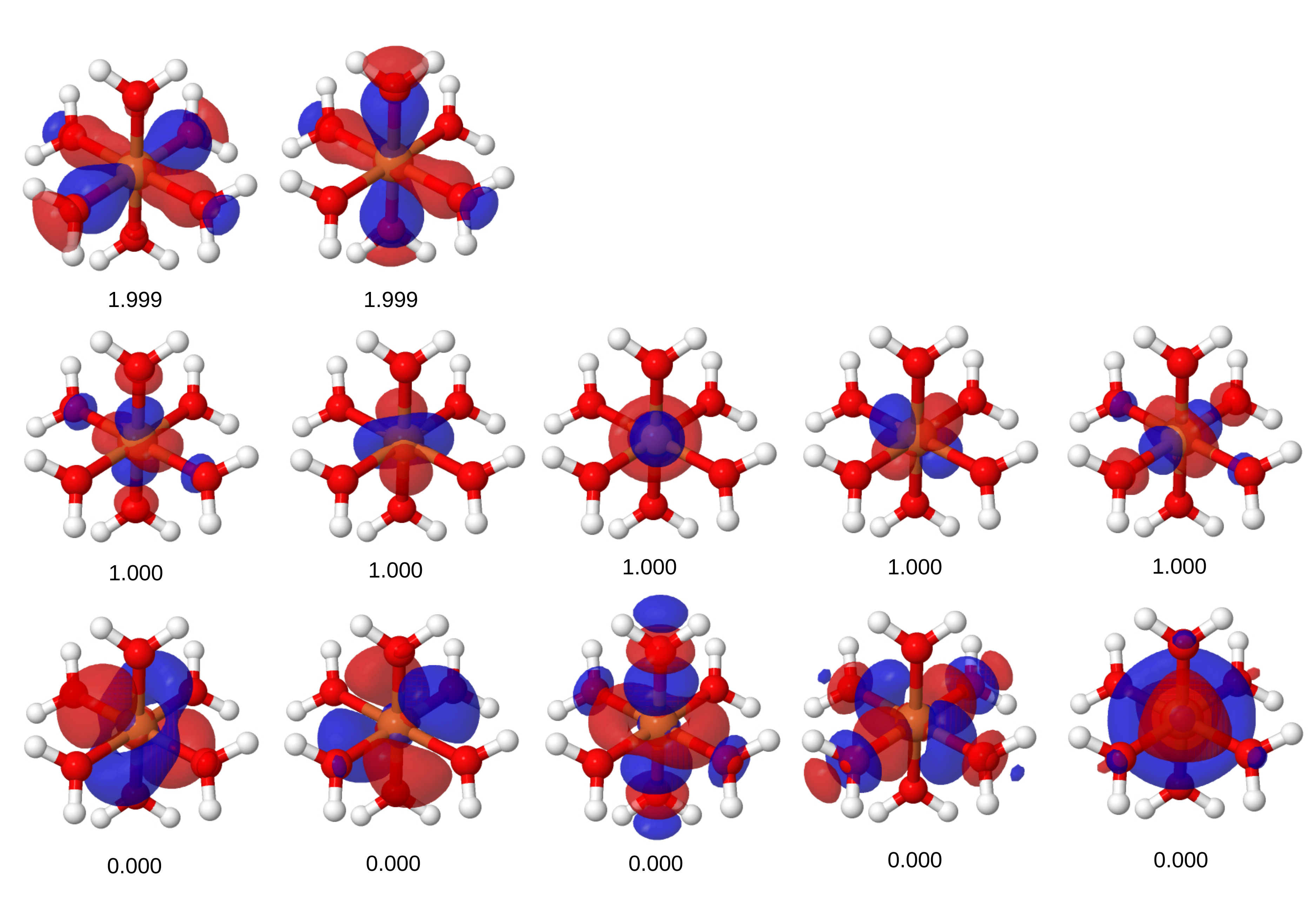}
    \caption{ Active orbitals for the sextet state of \ce{[Fe(H2O)6]^{3+}}. Occupation numbers are shown below (a similar active space was employed for the quartet state).\label{orbitals_2}}
    
\end{figure}

For \ce{[Fe(H2O)6]^{3+}} we focus on the $^6$A$_{g}$ and $^4$T$_{1g}$ lowest-energy states (using a nomenclature that assumes octahedral symmetry), and we used the structure from Ref.~\citenum{radon2016}, 
which was optimized for the $^6 \text{A}_{g}$ ground state within 
a cluster including the second solvation shell. We have removed this second solvation shell, and our calculations only concern the \ce{[Fe(H2O)6]^{3+}} core.
 Note that the CASPT2 value from Ref.~\citenum{radon2016} to which we compare was  obtained exactly this way (cf. Table 7 of Ref.~\citenum{radon2016}). It was shown that such a calculation includes a large  part of the solvation effect.   
For this complex, we employed the def2-TZVPP basis set\cite{weigend2005} for iron 
and oxygen, and the def2-SV(P) basis set for hydrogen.  Following Ref.~\citenum{radon2016} we employed an active space that includes the five $d$ orbitals, two additional $\sigma$ Fe-O bonding orbitals, and a second shell of $d$ orbitals yielding 
a CAS(9,12) active space. The orbitals comprising the active space are shown in Figure \ref{orbitals_2} 

As short-range exchange-correlation functionals we employed the spin-dependent srLDA functional from 
Ref.~\citenum{paziani2006} and the spin-dependent gradient-corrected functional of Ref.~\citenum{goll2006}, denoted by srPBE. The name of the latter reflects that it originates from the Perdew-Burke-Ernzerhof (PBE) functional\cite{perdew1996b}, and have been extended to the short-range interaction in Ref.~\citenum{TouColSav-JCP-05}, and further modified by Goll \textit{et al.}~\cite{goll2005,goll2006}.

\section{Demonstration of the open-shell CAS--srDFT method}\label{results}


We here discuss the results from CAS-srLDA and CAS-srPBE calculations on two prototype open-shell systems.

\subsection{The dioxygen molecule}

For dioxygen we start with a discussion of the optimal value of the range-separation parameter $\mu$. This matter has been discussed for CAS--srDFT\cite{fromager2007,hubert2016a} and also for other range-separated methods. 
Note that with an exact srDFT functional and a full CI long-range wave function (in a complete basis) the exact result would be obtained independent of the $\mu$ value. The $\mu$ value determines the mixture between the wave function and srDFT parts, and is thus a parameter that shifts correlation effects between the two parts.
When we refer to an ``optimal $\mu$ value" in the following, we mean the value giving accurate results with the least possible computational effort.
As the computational time for the srDFT part is basically independent of $\mu$ while the time for accurate long-range wave function contributions grows rapidly with $\mu$, the optimal $\mu$ value is the smallest $\mu$ value 
for which the non-local correlations not treated properly by semi-local srDFT functionals are \emph{not} in the realm of the srDFT part.
For multireference models, one  recipe has been to compare HF--srDFT and CAS--srDFT. From this, it was found that a value of $\mu$ of 0.4 bohr$^{-1}$ allocates the main part of the dynamical correlation to 
the srDFT functional, while the long-range wave function still incorporates a substantial part of static correlation in multiconfigurational systems.\cite{fromager2007}  The $\mu=0.4$ value 
was recently confirmed for excitation energies obtained from linear-response theory.\cite{hubert2016a}  
Here, we investigate the optimal value of $\mu$ for the \ce{O2} molecule in its triplet ground state $^3\Sigma^{-}_{g}$.
It is expected that this state is dominated by the electronic configuration
\begin{align}
\vert \Psi^{\ce{O2}}_{R_\text{eq}}(^3\Sigma^{-}_g)\rangle &= C_0 \; ^3 \vert 1\sigma^{2}_{g} 1\sigma^{*2}_{u} 2\sigma^{2}_{g} 2\sigma^{*2}_{u} 3\sigma^{2}_{g}1\pi^{2}_{u,x}1\pi^{2}_{u,y}1\pi^{*1}_{g,x}1\pi^{*1}_{g,y} 3\sigma^{*0}_{u} \rangle  + \cdots \nonumber \\
&\equiv C_0 \; ^3 \vert \pi^{2}_{u,x}\pi^{2}_{u,y}\pi^{*1}_{g,x}\pi^{*1}_{g,y} \rangle +\cdots
\end{align}
at the equilibrium bond distance.
The $\, ^{3}\vert\ldots\rangle$ notation includes the spin multiplicity of the configuration state function,
and we have in the last equality left out the $\sigma$ electrons for brevity. 
This dominant configuration can be described by a single-determinant high-spin wave function ($M_S = 1$), which means that HF--srDFT and CAS--srDFT should 
provide similar energies. In order to have a system that displays multiconfigurational character, and is directly comparable to the above system, we also 
consider a bond distance corresponding to a stretched bond ($R_{\text{stretch}} = 2.0$~\AA). In this case, we expect a splitting of HF--srDFT 
and CAS--srDFT energies, and the parameter $\mu$ should be chosen to reflect this. 

\begin{figure}
    \includegraphics[scale=0.40]{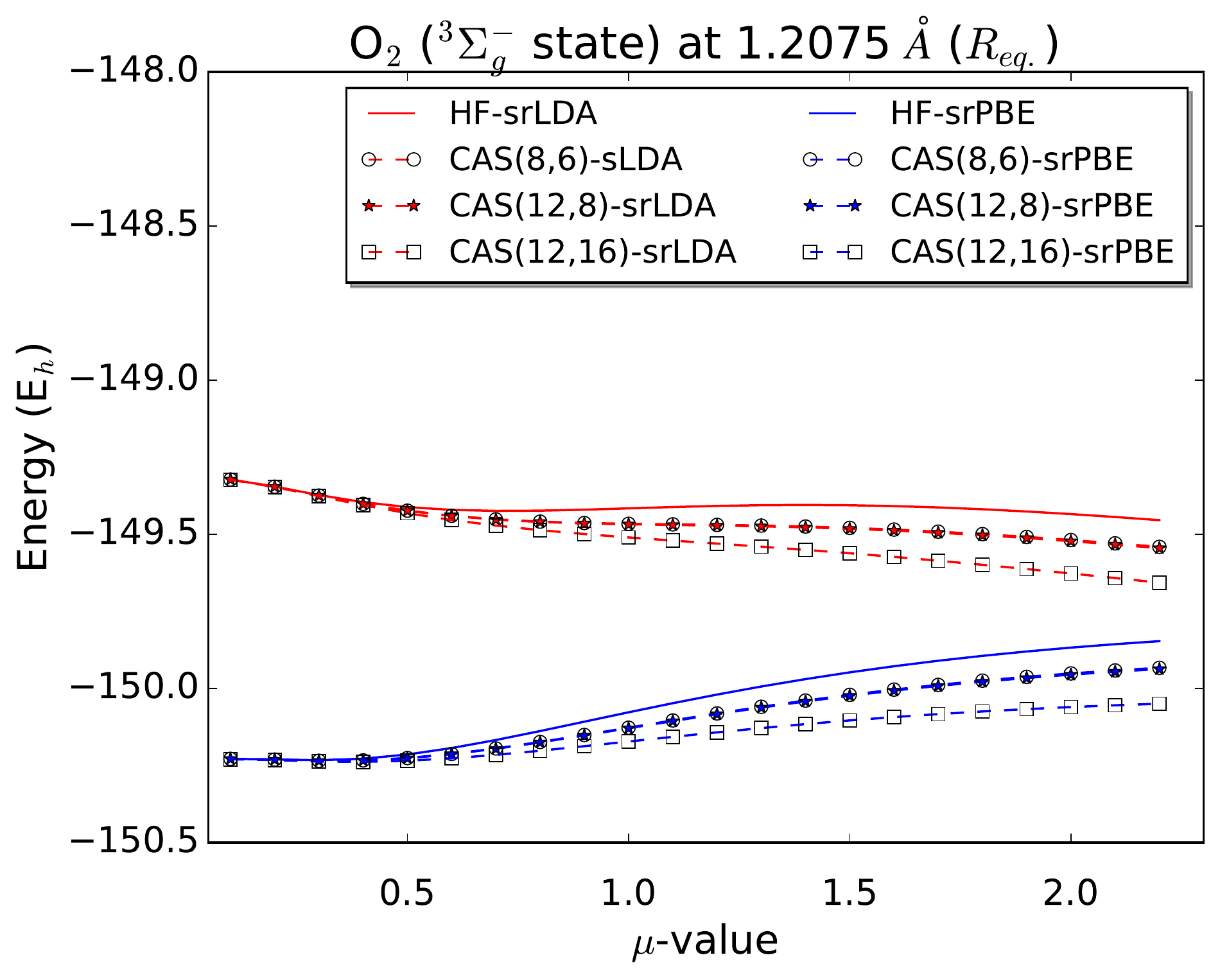}
        \includegraphics[scale=0.40]{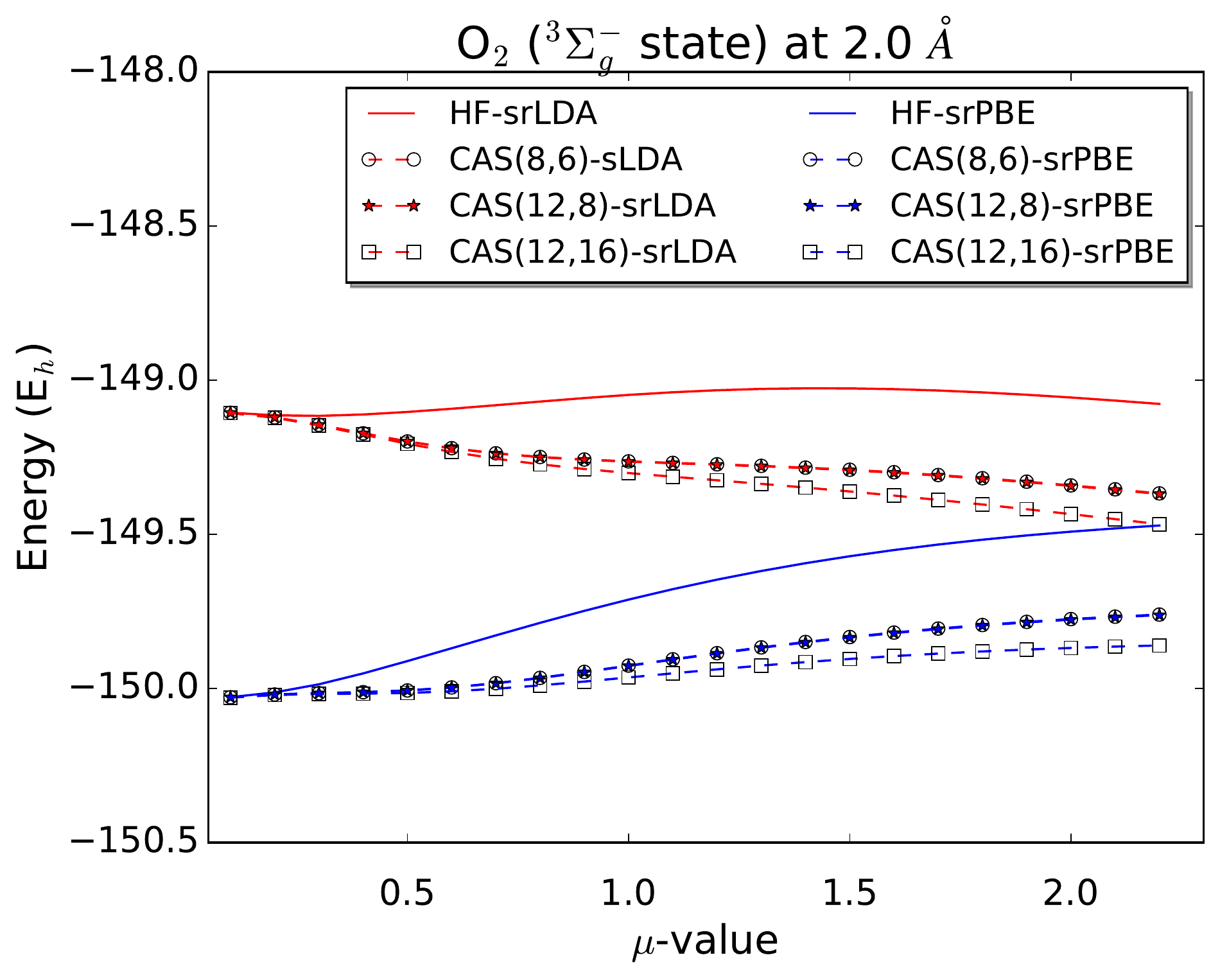}
    \caption{Dependence on the range-separation parameter $\mu$ of the HF--srDFT and CAS--srDFT energies for \ce{O2} in its $^3 \Sigma^{-}_{g} $ ground state for the equilibrium bond distance $R_\text{eq}$ (left) and a stretched bond distance $R_\text{stretch}$ (right). All calculations are with the cc-pVTZ basis set.}
    \label{mu-depend}
\end{figure}

The HF--srDFT and CAS--srDFT energies as functions of the value of $\mu$ are shown in Figure \ref{mu-depend}, both for $R_{\text{eq}}$ and $R_{\text{stretch}}$. For the equilibrium 
bond distance, the HF--srDFT and MC--srDFT energies are identical (within 10$^{-3}$ hartree) until we reach a value of $\mu \approx 0.4$.
This value is obtained for both of the employed srDFT functionals, and the result is thus very similar to what was found in Ref.~\citenum{fromager2007}.
We have tested three different active spaces of increasing size.
The smallest CAS(8,6) contains only $\sigma$ and $\pi$ orbitals from the oxygen 2p orbitals,
whereas the CAS(12,8) active space additionally includes the $\sigma$ orbitals with origin in the 2s orbitals and thus constitutes the full valence space.
Finally, the CAS(12,16) is an extended active space also including the oxygen 3s and 3p orbitals.
For $R_{\text{eq}}$ the splitting occurs at around the same value of $\mu$ (cf.~Figure \ref{mu-depend}, left), independent of the employed active space.    

For an elongated bond, the system becomes multiconfigurational as can be seen from the fact that the HF--srDFT and MC--srDFT energies are 
clearly different, already at small values of $\mu$ (cf.~Figure \ref{mu-depend}, right).  Thus, a value of $\mu=0.4$ also allows the long-range CAS wave function to include static correlation for the present multiconfigurational open-shell system.

We further comment on a few more observations from Figure \ref{mu-depend}. First, we note that for the srPBE functional, the $\mu$ dependence is benign for small $\mu$ values in case of \ce{O2}. For this small molecule the long-range CASSCF wave function can cover the correlation up to about $\mu=0.6-0.7$, as seen from the almost horizontal energy curves in this range of values of $\mu$.  As expected, this is in contrast with the CAS--srLDA energy, which is too high for $\mu=0$ and therefore decreases with increasing $\mu$, since the long-range wave function with increasing $\mu$ can describe an increasing part of the correlation. That this happens \textit{before} the HF-srLDA and MC-srLDA curves separate must mean that this effect is mostly in the exchange part. Further, that the CAS-srLDA energies decrease with increasing $\mu$ while CAS-srPBE energies increase indicates that pure CASSCF is better than any of the CAS-srLDA models, while CAS-srPBE contains some correlation effects not described by the pure CASSCF.

\begin{figure}
    \includegraphics[scale=0.40]{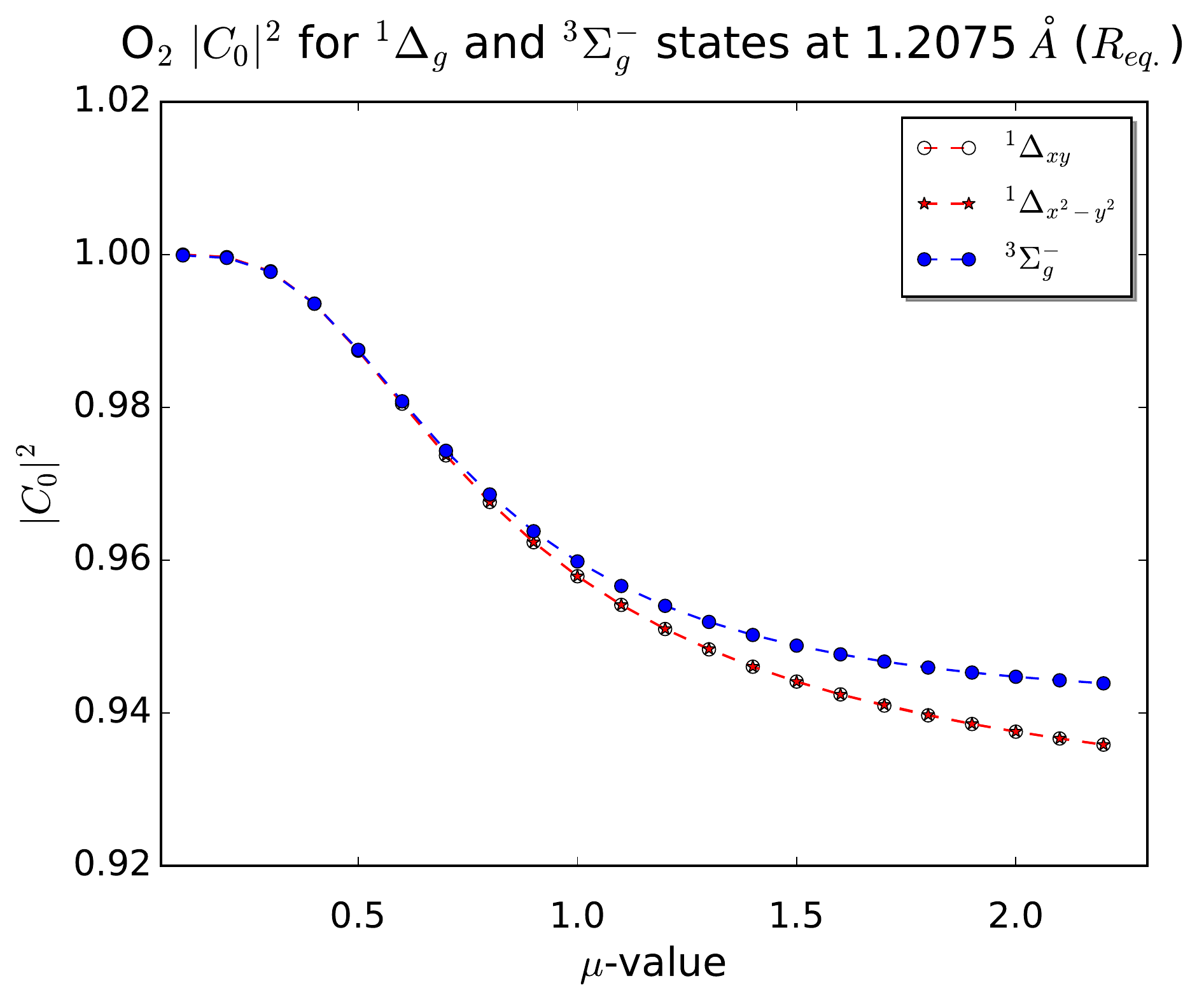}
        \includegraphics[scale=0.40]{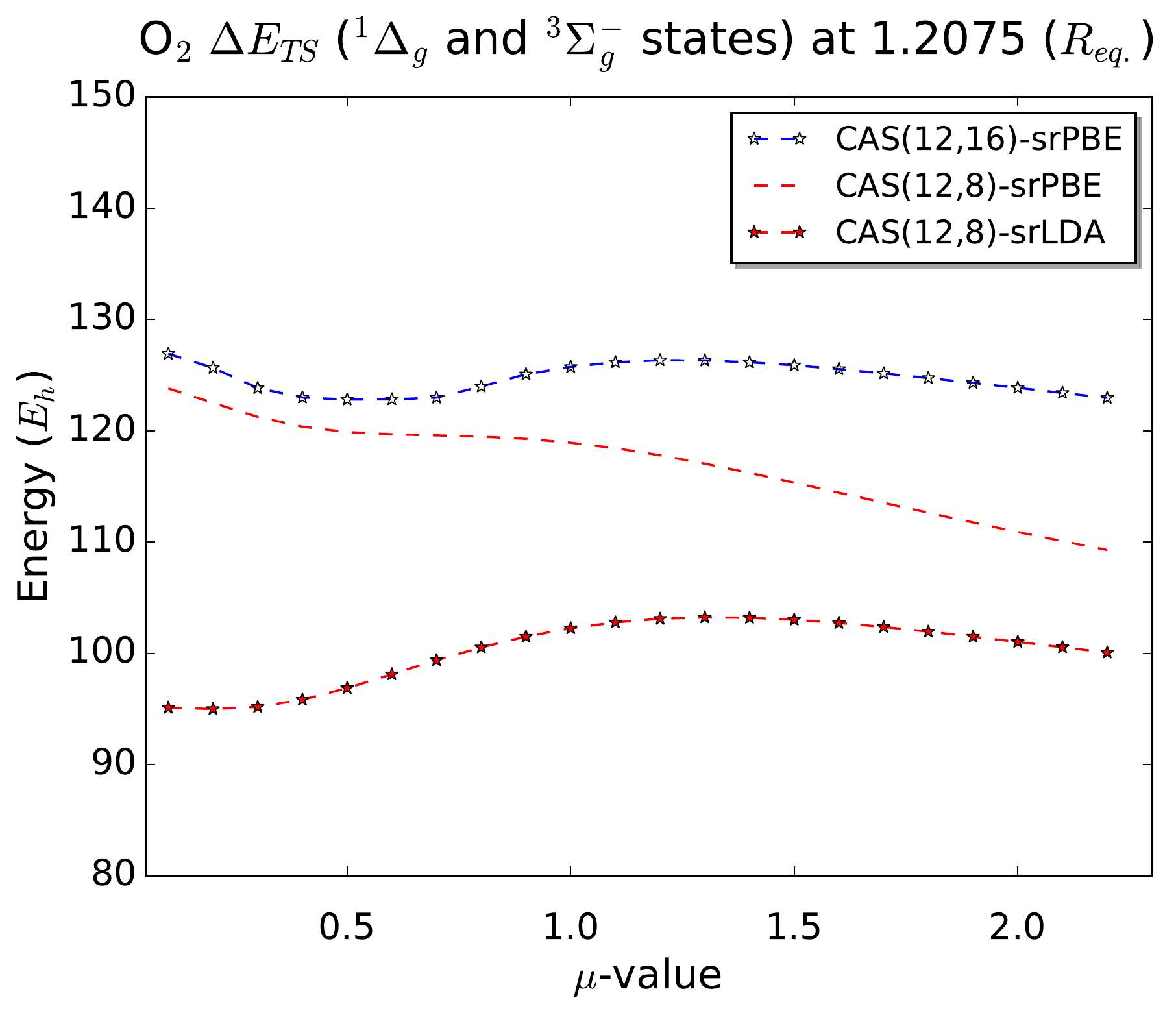}
    \caption{Left: Weight $|C_0|^2$ of the dominant configuration in the long-range CAS(12,8) wave function for \ce{O2} in the $^{3}\Sigma^{-}_{g}$ and $^{1} \Delta_{g}$ states at the equilibrium bond distance. The srLDA and srPBE functionals give rise to practically identical long-range wave functions. Right: Singlet-triplet splitting $\Delta E(^{3}\Sigma^{-}_{g} \rightarrow {}^{1}\Delta_g )$.}
    \label{singlet-triplet-o2}
\end{figure}

It is also interesting to study the first singlet excited state $^{1}\Delta_g$. In $\text{D}_{2h}$ symmetry this electronic state can either be characterized by a dominant configuration given by a linear combination of the Slater determinants in which the $\pi^{*}_{g}$ orbitals are doubly occupied, i.e. 
\begin{align}
\vert \Psi^{\ce{O2}}_{R_\text{eq}}(^1\Delta^{x^2-y^2}_{g})\rangle & =
C_0 \frac{1}{\sqrt{2}}(^1\vert  \pi^{2}_{u,x}\pi^{2}_{u,y}\pi^{*\,2}_{g,x}\pi^{*\,0}_{g,y}\rangle
 - {^1}\vert  \pi^{2}_{u,x}\pi^{2}_{u,y}\pi^{*\,0}_{g,x}\pi^{*\,2}_{g,y}\rangle) +\cdots
\label{delta_state_1} 
\end{align}
or by a dominant configuration in which the $\pi^{*}_{g,x}$ and $\pi^{*}_{g,y}$ orbitals are spin-singlet coupled
\begin{align}
 \vert \Psi^{\ce{O2}}_{R_\text{eq}}(^1\Delta^{xy}_{g})\rangle & =
 C_0 {\ }^{1}\vert  \pi^{2}_{u,x}\pi^{2}_{u,y}\pi^{*\,1}_{g,x}\pi^{*\,1}_{g,y}\rangle + \cdots.
\label{delta_state_2} 
\end{align} 
The states in Eqs.~\eqref{delta_state_1} and \eqref{delta_state_2} are exactly degenerate in $\text{D}_{\infty h}$ but belong to different irreducible representations in the $\text{D}_{2h}$
subgroup:
$\text{A}_{1g}$ and $\text{B}_{1g}$. None of the dominant configurations in these two wave functions can be represented by a spin-restricted single determinant, and thus cannot be obtained in standard spin-restricted Kohn-Sham DFT.
By contrast, the long-range wave function of MC--srDFT can describe the $^{1}\Delta_{g}$ state in both the $\text{D}_{2h}$ irreducible representations of Eqs.~\eqref{delta_state_1} and \eqref{delta_state_2}. The method correctly predicts that the $^1 \Delta^{x^2 - y^2 }_{g}$ and $^1 \Delta^{xy}_{g}$ states are degenerate. As expected, the dominant configurations in the long-range CAS wave functions are indeed given as the ones in Eqs.~\eqref{delta_state_1} and \eqref{delta_state_2}. In Figure \ref{singlet-triplet-o2} (left), we have reported the value of $|C_0 |^2$ for the dominant configuration in the long-range wave function as a function of $\mu$. 
The $^{3}\Sigma^{-}_{g}$, $^1 \Delta^{x^2 - y^2 }_{g}$, and $^1 \Delta^{xy}_{g}$ states behave remarkably similarly, in agreement with textbook qualitative MO theory, and the same trend is observed: A steadily decreasing value of $|C_0|^2$ as we increase $\mu$, reflecting that the long-range wave function must become more multiconfigurational to treat the electron correlation effects moved from the short-range to the long-range parts with increasing $\mu$. The value of $|C_0|^2$ for the $^1 \Delta$ states decreases fastest with increasing mixture of long-range wave function. Yet, at small values of $\mu$ (including $\mu=0.4$) we can expect that the singlet-triplet splitting will be determined to a large degree by the leading configuration for each state, and by how the short-range functionals translate the density obtained from the leading configuration into an energy. Any possible error is therefore likely due to the srDFT functional (as will be further discussed below). As a remark, we note that the $^1 \Delta^{x^2 - y^2 }_{g}$ and $^1 \Delta^{xy}_{g}$ energies as well as their respective $|C_0 |^2$ values are indistinguishable at all values of $\mu$.
This shows that the method treats singlet-coupled open shells qualitatively correct. 

With these results in mind, we can also investigate the singlet-triplet splitting 
$\Delta E(^3\Sigma^{-}_{g} \rightarrow {^{1}}\Delta_g )$. We note however, that we cannot expect accurate results due to deficiencies in the employed short-range functionals. 
 In particular, the short-range exchange does not correctly cancel the short-range Hartree self-repulsion for the singly occupied $\pi$ orbitals in singlet states.  In case of \ce{O2}, this can be seen from the above analysis of the dominant configurations: the long-range wave function correctly describes a spin-singlet coupled state in which the total spin density is zero but which locally may become both positive and negative (perhaps as shown most obviously in the $^1\Delta^{xy}_{g}$ state, Eq.~\eqref{delta_state_2}). Yet, our present srDFT functionals are not equipped with the ability to employ local spin densities, as required for a spin-singlet coupled state.  We therefore expect too high energy of the $^1\Delta_{g}$ state and thus of the splitting (as indeed observed, see below). Our investigation of the singlet-triplet splitting therefore focus mainly on the $\mu$-dependence whereas the values itself is only briefly discussed. 
The dependence on $\mu$ of the singlet-triplet splitting is shown in Figure \ref{singlet-triplet-o2} (right) for different active spaces and short-range functionals and given in Table \ref{splitting-energy-o2} for for selected values of $\mu$ with CAS(12,8)--srDFT and CAS(12,16)--srDFT. In general, the dependence on $\mu$ is moderate (at most a few kJ/mol) in the interval $\mu=0.1$--$0.5$.
The difference is larger ($\approx$ 25--30 kJ/mol) 
between the results obtained with the srLDA and srPBE functionals; thus there is a significant effect of employing a gradient-corrected functional.  Although the srLDA results are in fact closer to the experimental value than the srPBE results, both functionals have the above-mentioned flaws leading to destabilization of the singlet energy. Therefore, the good performance of srLDA must be considered fortuitous. We compare the splitting obtained with srLDA and srPBE to other results from the literature  in Table \ref{splitting-energy-o2_compare}. 
\begin{table}[t]
\caption{Vertical singlet-triplet splitting,
$\Delta E(^{3}\Sigma^{-}_{g} \rightarrow {}^{1}\Delta_g )$ (in kJ/mol),
for \ce{O2} at the equilibrium bond distance, calculated with CAS--srDFT employing different srDFT functionals and values of $\mu$ (in bohr$^{-1}$). \label{splitting-energy-o2}}
{ \footnotesize 
 \begin{tabular}{lcccccccc}
\toprule  \\[-1.0ex]
 \textbf{Method}        & ~$\mu=0.0$~ & ~$\mu=0.1$~  & ~$\mu = 0.3$~ & ~$\mu=0.4$~  & ~$\mu=0.5$~   & ~$\mu=1.0$~  & ~$\mu\to\infty$~ \\[1.5ex] 
\hline \\[-1.5ex]
CAS(12,8)-srLDA         &   96.96     &   95.11      &   95.18       &   95.82      &  96.86        &  102.24    &  92.35     \\[1.5ex]
CAS(12,16)-srLDA        &   96.96     &   96.97      &   97.66       &   100.13     & 103.53        &  107.78    &  110.44  \\[1.5ex]
CAS(12,8)-srPBE         &  127.62     &  123.79      &  121.24       &   120.36     & 119.88        &  118.92    &  92.75   \\[1.5ex]
CAS(12,16)-srPBE        &  127.62     &  126.91      &  123.83       &   122.99     & 122.80        &  125.74    &  110.44    \\[1.5ex]
\hline \\[-2.0ex]
Exp\cite{herzberg-diatomic}  & \multicolumn{7}{c}{ 94.6}  \\[1.0ex]
\hline\hline
 \end{tabular}
} 
\end{table}
\begin{table}[htb!]
\caption{Singlet-triplet splitting, $\Delta E(^{3}\Sigma^{-}_{g} \rightarrow {}^{1}\Delta_g )$ (in kJ/mol),
for \ce{O2} at the equilibrium bond distance.
The CAS-srLDA and CAS-srPBE results are for $\mu = 0.4$ bohr$^{-1}$.
Note that Refs.\citenum{gadzhiev2013,zen2014} report the adiabatic
energy difference, but with a difference of 0.008~\AA~between $R_e$ for the two states, this is expected to be of minor importance. \label{splitting-energy-o2_compare} }
 \begin{tabular}{lccccccc}
\toprule  \\[-1.0ex]
 \textbf{Method}             &     $\Delta E(^{3}\Sigma^{-}_{g} \rightarrow {}^{1}\Delta_g )$     \\[1.5ex] 
\hline \\[-1.5ex]
{\it Our work (all with cc-pVTZ):} &  \\[1.5ex]
CAS(12,8)                    &  92.35       \\[1.5ex]
CAS(12,8)-srLDA              &  95.82       \\[1.5ex]
CAS(12,8)-srPBE              &  120.36      \\[1.5ex]
CAS(12,16)                   &  110.44      \\[1.5ex]
CAS(12,16)-srLDA             &  100.13      \\[1.5ex]
CAS(12,16)-srPBE             &  122.99      \\[1.5ex] \hline \\[-1.5ex]
{\it Literature:} &  \\[1.5ex]
CCSD(T)/cc-pVTZ\cite{gadzhiev2013} & 124.66 \\[1.5ex] 
CCSDT/cc-pVTZ\cite{gadzhiev2013}   & 114.34 \\[1.5ex]
CCSDTQ/cc-pVTZ\cite{gadzhiev2013}  & 99.48 \\[1.5ex]
MRCI\cite{klotzchristel1984} & 111.61 \\[1.5ex] 
MRCI/cc-pVTZ\cite{gadzhiev2013} & 98.40 \\[1.5ex] 
QMC\cite{zen2014}            & 96.58 \\[1.5ex] \hline \\[-1.5ex]
Exp.\cite{herzberg-diatomic} & 94.6  \\[1.5ex] 
\hline \hline
 \end{tabular} 
\end{table} 
Our best theoretical CAS(12,16)--srPBE result yields a splitting of 123 kJ/mol, which is in between the CCSD(T) and CCSDT results, but (as expected) too large compared to the experimental value of 94.6 kJ/mol\cite{herzberg-diatomic}. The fact that
even CCSDT and some of the MRCI results\cite{klotzchristel1984} overestimate the experiment value by 20 and 16 kJ/mol, respectively, emphasizes that the calculation of this singlet-triplet splitting energy is indeed delicate.
 Only the highly accurate CCSDTQ and QMC methods are able to achieve a result that is within 5 kJ/mol of the experimental value. 
A peculiar observation is that the small active space CAS(12,8) is very close to the experimental result, and CAS(12,8)-srLDA is almost identical to experiment. Both results are due to fortuitous error cancellation, 
as extending the active space leads to deterioration. Yet, CAS(12,16) and CAS(12,16)-srLDA are in fact still closer to experiment than both CCSDT and MRCI, showing that assessment of the functionals can require some consideration of the active space employed. Finally, we note that approaches 
based on perturbation theory (e.g.~CASPT2 or NEVPT2) are likely also to provide a reasonable spin-state splitting. Indeed, we performed a NEVPT2(12,8) calculation which gives a singlet-triplet splitting of 92.35 kJ/mol. However, an in-depth discussion of CASPT2/NEVPT2 results is out of the scope for our current paper, where we focus on MC--srDFT. To improve spin-state energetics, we are planning to make a number of improvements on the srDFT functionals, e.g., employing local spin densities. One option could be to employ the on-top pair density\cite{perdew1995,manni2014} for this purpose.

\subsection{The $^{6}\text{A}_{g}\rightarrow{}^{4}\text{T}_{2g}$ spin-state splitting in \ce{[Fe(H2O)6]^{3+}}  }

We now turn to a system where methods such as CCSDTQ are probably beyond reach. The prototypical \ce{[Fe(H2O)6]^{3+}} complex has been described in numerous theoretical works with a variety of methods. We will here focus on the $^{6}\text{A}_{g}\rightarrow{}^{4}\text{T}_{2g}$ transition, where in our labeling we assume octahedral symmetry. The higher spin state ($^{6}\text{A}_{g}$) can qualitatively be described as an open-shell d-complex with five singly occupied d-orbitals, whereas the lower spin state ($^{4}\text{T}_{2g}$) can be described as a doubly occupied orbital in the lowest-lying $t_{2g}$ orbital level and a hole in the $e_g$ level. The lower spin-state thus does not involve a coupling between spatially separated orbitals, suggesting that we will not observe errors due to the lack of local spin dependence in the employed srDFT functionals as described in the previous section. 

The experimental band for the $^{6}\text{A}_{g}\rightarrow{}^{4}\text{T}_{2g}$ transition has a maximum at 12600 cm$^{-1}$ (1.56 eV) in acidic, aqueous solution\cite{jorgensen1954}. 
However, previous results obtained with CASPT2 and CCSD(T) have showed remarkably large deviations (up to 1.5 eV) from this value\cite{ghosh2003,yang2014}. Large deviations 
were also found with the spectroscopically oriented configuration interaction (SORCI) method\cite{neese2007}. 
On the other hand, both
DFT\cite{harris1997,hughes2011,yang2013} and the semi-empirical INDO/S method seem to be reasonably accurate\cite{anderson1986,harris1997}.  
A recent study by Rado{\'n} \textit{et al.}\cite{radon2016} pointed out some of the possible failures of accurate \textit{ab initio} methods. First, it was argued that 
the second solvation shell was important, mainly for the molecular structure of the \ce{[Fe(H2O)6]^{3+}} core.
If this effect is taken into account, even calculations on \ce{[Fe(H2O)6]^{3+}} 
are in significant better agreement with experiment. Further, Rado{\'n} \textit{et al.}\cite{radon2016} argued that in addition to the $d$ orbitals, 
the employed CAS space should at least contain two 
$\sigma$ Fe-O bonding orbitals.\cite{yang2014} This leads  
to a CAS space of 9 electrons in 12 orbitals, whereas some earlier studies employed the smaller CAS(5,5) and CAS(5,10) active spaces.\cite{yang2014}. The effect was estimated 
to be around 0.24 eV on the transition energy.
However, Rado{\'n} \textit{et al.}\cite{radon2016} also stressed that their study was not directly comparable to some of the previous studies as they 
(unlike  Ref.~\citenum{yang2014}) employed an IPEA-shifted zeroth-order Hamiltonian
and also larger basis sets of which the combined effect can be estimated to be around 0.3 eV on the transition energy.

Here, we start by investigating the dependence on $\mu$ of the spin-state splitting $\Delta E(^6\text{A}_{g}\rightarrow{{}^{4}\text{T}_{1g}})$. 
Results for selected values of $\mu$ are shown in Table \ref{splitting-energy-feh2o63+}. Compared to the spin-state splitting in \ce{O2} (see previous section), 
the $\Delta E(^6\text{A}_{g}\rightarrow{}^{4}\text{T}_{1g})$ spin-state splitting in the range $\mu = 0.4-1.0$ bohr$^{-1}$ 
varies somewhat more with 
1.15--2.10 eV for CAS(9,12)-srPBE. For values in the interval $\mu=0.33$ to $\mu=0.50$ bohr$^{-1}$ expected \emph{a priori} to be optimal, we do obtain good agreement with the experimental value of 1.56 eV. The "end-points" in Table \ref{splitting-energy-feh2o63+} corresponds to a pure CAS(9,12) ($\mu=\infty$) and pure PBE ($\mu=0.01$); these respectively overestimate and underestimate the experimental value significantly. It is gratifying the range of $\mu$-values expected to provide the best results, indeed all are the best ones obtained (an in fact, all closer to experiment than CASPT2 with srPBE). Yet, before more general conclusions concerning the performance can be made, a larger span of systems must be explored. In Table \ref{splitting-energy-feh2o6_compare}, we compare the CAS(9,12)-srPBE result at $\mu=0.4$  bohr$^{-1}$ with the
CAS(9,12)-srLDA one, as well as results obtained with a number of other theoretical methods.       
\begin{table}[t]
\small
\caption{Vertical quartet-sextet splitting (in eV) of \ce{[Fe(H2O)6]^{3+}}, calculated with CAS--srPBE employing 
	different values of $\mu$ (in bohr$^{-1}$). 
\label{splitting-energy-feh2o63+} }
 \begin{tabular}{lcccccccc}
\toprule  \\[-1.0ex]
 \textbf{Method}   & ~$\mu=0.01$~ & ~$\mu=0.1$~  &  ~$\mu=0.33$~ & ~$\mu=0.4$~ & ~$\mu=0.5$~ & $~\mu = 0.75$~ & ~$\mu = 1.0$~ & ~$\mu=\infty$~  \\[1.5ex] 
\hline \\[-1.5ex]
 CAS(9,12)-srPBE   &  1.15      & 1.15       &    1.35     &  1.44     &    1.61   &  1.74        &  2.10         &  2.24       \\[1.5ex]
\hline \hline
 \end{tabular} 
\end{table}
As mentioned above, the literature results show a rather large scatter, and all the values for the quartet-sextet splitting are too large.
Our result with CAS(9,12)-srLDA shows the opposite pattern, it underestimates the experimental value significantly. The corresponding values for $\mu=0.33$ and $\mu=0.5$ are 1.04 eV and 1.29 eV and are thus still significantly underestimated. 
However, the gradient-corrected srDFT functional significantly improves the spin-state splitting energy for the value $\mu=0.4$. 
This shows that a GGA srDFT functional is important, as it also was for \ce{O2}.
We found in previous section that $\mu=0.4$ was a good compromise between flexibility of the long-range wave function to include static correlation and ability of the srDFT functional to recover dynamical correlation. 
This value also provides a very accurate estimate for the quartet-sextet splitting $\Delta E(^6\text{A}_{g}\rightarrow{}^{4}\text{T}_{1g})$ with CAS(12,9)-srPBE. The result is 
1.44 eV, only 0.12 eV below the experimental value.
In comparison, a calculation with CAS(9,12)PT2 on the same \ce{[Fe(H2O)6]^{3+}} molecular structure gives a spin-state splitting of 1.96 eV, 
which is 0.44 eV too high. 
We note that $\mu=0.33$ and $\mu=0.5$ also give splittings better than the CAS(9,12)PT2 values.
\begin{table}
\caption{Vertical quartet-sextet splitting (in eV) for \ce{[Fe(H2O)6]^{3+}}. The CASPT2, CAS(9,12)-srLDA, and CAS(9,12)-srPBE calculations are for the \ce{[Fe(H2O)6]^{3+}} core with a molecular structure 
that included the second solvation shell in the optimization, see Ref.~\citenum{radon2016}. The CAS(9,12)-srLDA and CAS(9,12)-srPBE calculations are for a range-separation parameter of $\mu=0.4$ bohr$^{-1}$.
 \label{splitting-energy-feh2o6_compare} }
    \begin{tabular}{lcccc} 
      \hline \hline \\[-1.5ex]
       \textbf{Method}                 & ~~~~ $\Delta E(^{6}\text{A}_{g}\rightarrow{}^{4}T_{1g})$  \\[0.5ex]
\hline \\[-1.5ex]
       \quad \ce{[Fe(H2O)6]^{3+}} optimized in vacuum & \\[1.5ex]  
       {\it Literature:} &  \\[1.5ex]  
       CCSD(T)\cite{ghosh2003}         &  ~~~~2.47       \\[0.5ex]
       SORCI\cite{neese2007}           &  ~~~~2.07       \\[0.5ex]
       CASPT2\cite{ghosh2003}          &  ~~~~2.64       \\[0.5ex]
              \hline \\[-1.5ex]
       \quad \ce{[Fe(H2O)6]^{3+}} optimized with a second solvation sphere &  \\[1.5ex]
       {\it Literature:} &  \\[1.5ex]
       CASPT2\cite{radon2016}          &  ~~~~1.96       \\[1.5ex]
       {\it Our work:} &  \\[1.5ex]
       CAS(9,12)-srLDA                 &  ~~~~1.14       \\[0.5ex] 
       CAS(9,12)-srPBE                 &  ~~~~1.44       \\[0.5ex]
\hline \\[-2.0ex]
       Exp.\cite{jorgensen1954}        &  ~~~~1.56       \\[0.5ex]
      \hline \hline
    \end{tabular} 
    \\[1.0ex]
      
\end{table}

\section{Conclusions}\label{conclusion}

In this paper, we have presented the theory and implementation of  a generalization of the MC--srDFT method to employ spin-dependent short-range density functionals.
We have applied the method to two well-known cases, namely the \ce{O2} molecule which has a triplet ground state, and the transition metal complex \ce{[Fe(H2O)6]^{3+}}. 

For the \ce{O2} molecule, a wealth of literature data exist for the spin-state splitting between the $^3\Sigma^{-}_{g}$ ground state and the lowest-lying singlet state  $^{1}\Delta_g$, and QMC or CCSDTQ methods provide a value in very good agreement with experiment. 
The spin-state splitting 
 has here been investigated with a range of values for the range-separation parameter $\mu$, but the splitting is not strongly dependent on $\mu$. Compared to the experimental value,
CAS--srPBE lies between CCSD(T) and CCSDT in accuracy. This is already impressive, but there is also room for improvement, as the $^1 \Delta_{g}$ state is estimated too high in energy by both CCSD(T), CCSDT and CAS-srDFT. The former two methods can be improved by higher-order cluster expansions, but this comes at a high computational cost. An improvement (that will be computationally cheap) for CAS-srDFT is to allow dependence of local spin densities in the employed srDFT functionals, which is expected to lower the energy of the spin-coupled $^1 \Delta_{g}$ state.  

Next, we investigated the spin-state splitting $\Delta E(^{6}\text{A}_{g}\rightarrow{}^{4}\text{T}_{1g})$ in \ce{[Fe(H2O)6]^{3+}}.
In this case, CCSD(T) and CASPT2 are both significantly off. A method such as CCSDTQ is too computationally demanding, while CAS--srPBE shows very good agreement with experiment.
 
It should also be emphasized that most other high-level methods require large basis set  expansions with high-order angular momenta to describe the electron-electron Coulomb cusp accurately. The MC--srDFT method avoids this by replacing an explicit description of the Coulomb cusp with an effective density functional. Hence, the basis set requirements can be expected to be similar to regular DFT, which is known for its fast convergence with respect to basis set expansion. 

The first studies shown here are promising, but it should be emphasized that further developments are required. An obvious extension is to employ srDFT functionals that include the kinetic energy density, which has been developed for range-separated methods.\cite{goll2009}
Also, a good measure of local spin density, for example in terms of the on-top pair density~\cite{perdew1995}, needs to be implemented to
satisfactorily describe for example open-shell singlets which exhibit zero mean spin density.
Extension of the open-shell algorithm to molecular properties via response theory will also be an important extension of the method.  Developments in these directions are ongoing.     
   
In the future it would also be interesting to extend our state-averaged version of MC-srDFT\cite{hedegaard2015b}, in particular for unbiased location of conical intersections.
 
\appendix

\section{Gradient contributions from short-range GGA functionals} \label{app_GGA}

For an srGGA functional, the short-range exchange-correlation energy density $e_{\text{sr-xc}}$ also depends on the three gradient variables
$\xi_{CC}(\mathbf{r}) = \nabla\rho_C(\mathbf{r})\cdot\nabla\rho_C(\mathbf{r})$,
$\xi_{SS}(\mathbf{r}) = \nabla\rho_S(\mathbf{r})\cdot\nabla\rho_S(\mathbf{r})$, and
$\xi_{CS}(\mathbf{r}) = \nabla\rho_C(\mathbf{r})\cdot\nabla\rho_S(\mathbf{r})$
in addition to the electron charge density $\rho_C(\mathbf{r})$ and the electron spin density $\rho_S(\mathbf{r})$:
\begin{equation}
 E_{\text{sr-xc}}[\rho_{C}, \rho_S, \xi_{CC}, \xi_{SS},\xi_{CS}] =  \int e_{\text{sr-xc}}(\rho_{C}(\mathbf{r}),\rho_S(\mathbf{r}),\xi_{CC}(\mathbf{r}),\xi_{SS}(\mathbf{r}),\xi_{CS}(\mathbf{r}))\,\text{d}\mathbf{r}.  
\label{E_xc_GGA}
\end{equation}
where we for brevity define the integrand as $e_{\text{sr-xc}}$. 
The derivatives required for the electronic gradient then become 
\begin{align}
 \frac{\partial e_{\text{sr-xc}}}{\partial \lambda_i } & = 
 \frac{\partial e_{\text{sr-xc}}}{\partial\rho_{C}} \frac{\partial \rho_{C}(\mathbf{r})}{\partial \lambda_i} + 
 \frac{\partial e_{\text{sr-xc}}}{\partial\rho_{S}} \frac{\partial \rho_{S}(\mathbf{r})}{\partial \lambda_i} \notag \\
&+ \frac{\partial e_{\text{sr-xc}}}{\partial\xi_{CC}} \frac{\partial \xi_{CC}(\mathbf{r})}{\partial \lambda_i}
 + \frac{\partial e_{\text{sr-xc}}}{\partial\xi_{SS}} \frac{\partial \xi_{SS}(\mathbf{r})}{\partial \lambda_i}
 + \frac{\partial e_{\text{sr-xc}}}{\partial\xi_{CS}} \frac{\partial \xi_{CS}(\mathbf{r})}{\partial \lambda_i} \notag \\
\end{align}
and the derivatives required for the electronic Hessian become 
\begin{align}
  \frac{\partial^2 e_{\text{sr-xc}}}
  {\partial \lambda_i \partial \lambda_j} 
 & = 
 \sum_{X=C,S}\left(\frac{\partial e_{\text{sr-xc}}}{\partial\rho_{X}} \frac{\partial^{2} \rho_{X}(\mathbf{r})}{\partial \lambda_i \partial  \lambda_j} 
+ \frac{\partial e_{\text{sr-xc}}}{\partial\xi_{XX}} \frac{\partial^{2} \xi_{XX}(\mathbf{r})}{\partial \lambda_i \partial \lambda_j}\right) 
+ \frac{\partial e_{\text{sr-xc}}}{\partial\xi_{CS}} \frac{\partial^{2} \xi_{CS}(\mathbf{r})}{\partial \lambda_i \partial \lambda_j} \notag \\
& +  \sum_{X,Y=C,S}  \left(\frac{\partial^{2} e_{\text{sr-xc}}}{\partial\rho_{Y}\partial\rho_{X}}
\frac{\partial \rho_{X}(\mathbf{r})}{\partial \lambda_i}\frac{\partial \rho_{Y}(\mathbf{r})}{\partial \lambda_j} + 
\frac{\partial^{2} e_{\text{sr-xc}}}{\partial\rho_Y\partial\xi_{XX}}
\frac{\partial \xi_{XX}(\mathbf{r})}{\partial \lambda_i}\frac{\partial \rho_{Y}(\mathbf{r})}{\partial \lambda_j}  \right. \notag \\
 & +  
\left. \frac{\partial^{2} e_{\text{sr-xc}}}{\partial \xi_{YY} \partial \rho_{X}}
\frac{\partial \rho_{X}(\mathbf{r})}{\partial \lambda_i}\frac{\partial \xi_{YY}(\mathbf{r})}{\partial \lambda_j}  +  
 \frac{\partial^{2} e_{\text{sr-xc}}}{\partial \xi_{YY}\partial\xi_{XX}}
\frac{\partial \xi_{XX}(\mathbf{r})}{\partial \lambda_i}\frac{\partial \xi_{YY}(\mathbf{r})}{\partial \lambda_j} \right) \notag\\
&+ \sum_{X=C,S}\left(
\frac{\partial^{2} e_{\text{sr-xc}}}{\partial \xi_{CS} \partial \rho_{X}}
\frac{\partial \rho_{X}(\mathbf{r})}{\partial \lambda_i}\frac{\partial \xi_{CS}(\mathbf{r})}{\partial \lambda_j}  +  
\frac{\partial^{2} e_{\text{sr-xc}}}{\partial \xi_{CS}\partial\xi_{XX}}
\frac{\partial \xi_{XX}(\mathbf{r})}{\partial \lambda_i}\frac{\partial \xi_{CS}(\mathbf{r})}{\partial \lambda_j}
 \right) \notag \\
&+ \frac{\partial^{2} e_{\text{sr-xc}}}{\partial \xi_{CS}^2}
 \frac{\partial \xi_{CS}(\mathbf{r})}{\partial \lambda_i}\frac{\partial \xi_{CS}(\mathbf{r})}{\partial \lambda_j}.
\end{align}
The first-order derivatives of the three gradient terms are
\begin{align}
\frac{\partial \xi_{CC} (\mathbf{r})}{\partial \lambda_i } &= 2\sum_{pq} ( \nabla\Omega_{pq}(\mathbf{r})\cdot\nabla\rho_{C}(\mathbf{r}) ) \frac{\partial D^{C}_{pq}}{\partial \lambda_i }, \\
\frac{\partial \xi_{SS} (\mathbf{r})}{\partial \lambda_i } &= 2\sum_{pq} ( \nabla\Omega_{pq}(\mathbf{r})\cdot\nabla\rho_{S}(\mathbf{r}) ) \frac{\partial D^{S}_{pq}}{\partial \lambda_i }, \\
\frac{\partial \xi_{CS} (\mathbf{r})}{\partial \lambda_i } &= \sum_{pq} \nabla\Omega_{pq}(\mathbf{r})\cdot\left(
\nabla\rho_{C}(\mathbf{r}) \frac{\partial D^{S}_{pq}}{\partial \lambda_i } +
\nabla\rho_{S}(\mathbf{r})\frac{\partial D^{C}_{pq}}{\partial \lambda_i } \right),
\end{align}
and the second-order derivatives are
\begin{align}
\frac{\partial^2 \xi_{CC} (\mathbf{r})}{\partial \lambda_i \partial \lambda_j} &= 
2\sum_{pq,rs} ( \nabla\Omega_{pq}(\mathbf{r}) \cdot \nabla \Omega_{rs}(\mathbf{r}) )
\frac{\partial D^{C}_{rs}}{\partial \lambda_j } \frac{\partial D^{C}_{pq}}{\partial \lambda_i }
 \notag \\
&+2\sum_{pq} ( \nabla\Omega_{pq}(\mathbf{r}) \cdot \nabla\rho_{C}(\mathbf{r}) )
\frac{\partial^2 D^{C}_{pq}}{\partial \lambda_i \partial \lambda_j}, \\
\frac{\partial^2 \xi_{SS} (\mathbf{r})}{\partial \lambda_i \partial \lambda_j} &= 
2\sum_{pq,rs} ( \nabla\Omega_{pq}(\mathbf{r}) \cdot \nabla \Omega_{rs}(\mathbf{r}) )
\frac{\partial D^{S}_{rs}}{\partial \lambda_j } \frac{\partial D^{S}_{pq}}{\partial \lambda_i }
\notag \\
&+2\sum_{pq} ( \nabla\Omega_{pq}(\mathbf{r}) \cdot \nabla\rho_{S}(\mathbf{r}) )
\frac{\partial^2 D^{S}_{pq}}{\partial \lambda_i \partial \lambda_j}, \\
\frac{\partial^2 \xi_{CS} (\mathbf{r})}{\partial \lambda_i \partial \lambda_j} &= 
\sum_{pq,rs} ( \nabla\Omega_{pq}(\mathbf{r}) \cdot \nabla \Omega_{rs}(\mathbf{r}) )
\left( \frac{\partial D^{C}_{rs}}{\partial \lambda_j } \frac{\partial D^{S}_{pq}}{\partial \lambda_i } 
+ \frac{\partial D^{S}_{rs}}{\partial \lambda_j } \frac{\partial D^{C}_{pq}}{\partial \lambda_i } \right)
\notag \\
&+\sum_{pq} \nabla\Omega_{pq}(\mathbf{r}) \cdot \left(
   \nabla\rho_{C}(\mathbf{r}) \frac{\partial^2 D^{S}_{pq}}{\partial \lambda_i \partial \lambda_j}
  +\nabla\rho_{S}(\mathbf{r}) \frac{\partial^2 D^{C}_{pq}}{\partial \lambda_i \partial \lambda_j}
   \right) .
\end{align}

\begin{acknowledgments}

EDH thanks the Carlsberg foundation and the European Commission for post-doc stipends. JT and HJJ acknowledge support from the French government via the French-Danish collaboration project ``WADEMECOM.dk''.

\end{acknowledgments}

\bibliography{srdft}

\end{document}